\documentclass[a4paper,twocolumn,11pt,unpublished]{quantumarticle}
\pdfoutput=1
\usepackage[utf8]{inputenc}
\usepackage[english]{babel}
\usepackage[T1]{fontenc}
\usepackage{amsmath}
\usepackage{amssymb}
\usepackage{amsthm}
\usepackage{bm}
\usepackage{graphicx}
\usepackage{booktabs}
\usepackage[numbers,sort&compress]{natbib}
\usepackage{hyperref}

\newenvironment{ruledtabular}{\par\centering\small}{\par}

\begin{document}

\title{QDMin: Coherent Quantum Minimum Finding over Parametrized Many-Body Hamiltonians}

\author{Abhishek}
\orcid{0000-0003-2226-3146}
\email{abhishek@ph.iitr.ac.in}
\affiliation{Department of Physics, Indian Institute of Technology Roorkee, Roorkee 247667, India}

\author{P. Arumugam}
\orcid{0000-0001-9624-8024}
\email{arumugam@ph.iitr.ac.in}
\affiliation{Department of Physics, Indian Institute of Technology Roorkee, Roorkee 247667, India}
\affiliation{Centre for Photonics and Quantum Communication Technology, Indian Institute of Technology Roorkee, Roorkee 247667, India}

\maketitle

\begin{abstract}
Many questions about interacting quantum systems ask which member of a
Hamiltonian family \(H(\bm q)\) has the lowest ground-state energy, rather
than for the ground state of one Hamiltonian. A pointwise scan answers this
with one many-body solve per candidate. QDMin (Quantum Deformation Minimum
Finding) instead holds all \(N_q\) candidates in a coherent register and
searches them with a ground-energy threshold oracle, replacing \(N_q\)
solves by \(O(\sqrt{N_q})\) coherent oracle calls. We identify static-field
(Hubbard--Stratonovich) families as a broad class in which this minimum is a
rigorous variational bound that the parent ground state does not reveal, and
we make the oracle robust to imperfect ground-state preparation by
fixed-point amplitude amplification. The measured search constant places the
crossover against a pointwise quantum scan at \(N_q\approx220\,\xi^2\), where
\(\xi\) is the overhead of a coherent call. We test the search on realistic
\textit{sd}-shell (USDB) static quadrupole landscapes with up to \(10^6\)
grid points and on a Hubbard ring with competing orders. It pays where
competing minima trap local descent, needing about four times fewer energy
queries than multistart descent for oblate--prolate coexistence in
\(^{28}\)Si, and not on smooth single-well landscapes. Circuit-level
simulations validate the complete pipeline on a schematic model. The gain
concerns the number of energy queries; any advantage of the quantum solver
itself is separate.
\end{abstract}

\section{Introduction}

In many problems the question is not the ground state of one Hamiltonian
but which member of a family \(H(\bm q)\) has the lowest ground-state energy.
The parameters \(\bm q\) may be static
order-parameter fields, collective shape coordinates, or external controls,
and the target is \(\bm q^\star=\arg\min_{\bm q}E_0(\bm q)\) with
\(E_0(\bm q)=\min\operatorname{spec}H(\bm q)\). When each \(E_0(\bm q)\)
requires a quantum many-body solver, the direct approach is a pointwise scan:
solve at each of \(N_q\) candidate points and keep the smallest result. All
of the cost lies in the \(N_q\) solves. This paper shows how to locate
\(\bm q^\star\) with \(O(\sqrt{N_q})\) coherent oracle calls instead, each
costing a fixed number of solve-equivalents, without constructing the
landscape, and quantifies when that reduction pays: for large,
multidimensional parameter grids with competing minima, but not for short or
smooth scans.

In nuclear structure, \(\bm q\) are multipole deformations: mean-field and density-functional
calculations organize intrinsic states over quadrupole and higher multipole
coordinates~\cite{bender2003selfconsistent}, competing minima underlie shape
coexistence~\cite{heydewood2011coexistence}, and octupole and fission
degrees of freedom make the landscape multidimensional
~\cite{butler2016octupole,schunckrobledo2016fission,
xiang2026triaxialquadrupoleoctupole}. Constrained Hartree--Fock--Bogoliubov
solvers such as HFODD and HFBTHO scan these coordinates systematically
~\cite{dobaczewski2021hfodd,marevic2022hfbtho}, and the cost of such scans has
motivated emulation~\cite{lay2024fissionemulation}.

A broad and physically important class of such families arises from static
Hubbard--Stratonovich decoupling~\cite{stratonovich1957method,hubbard1959partition}
of attractive quadratic interactions. If
\(H=H_0-\sum_a(\chi_a/2)A_a^2\), the static-field family
\(H(\bm q)=H_0-\sum_a\chi_aq_aA_a+\sum_a(\chi_a/2)q_a^2\) satisfies
\(H(\bm q)\succeq H\) for every \(\bm q\)
(Sec.~\ref{sec:problem}). Its minimum is therefore the best static-field
variational bound, its minimizer is the self-consistent order parameter, and
all correlations outside the decoupled channels are treated exactly.
Finding \(\bm q^\star\) is not the same as finding the ground state of \(H\):
a symmetric ground state, such as the \(J=0\) state of a deformed nucleus,
has vanishing order parameter and does not reveal \(\bm q^\star\).

QDMin (Quantum Deformation Minimum Finding) reduces the number of solves by
promoting \(\bm q\) to a coherent register. The family is implemented as a
single operator \(H(\hat{\bm q})\), whose cost grows polynomially in the
register size rather than with \(N_q\). A coherent oracle prepares the
ground-state component on every parameter branch, estimates its energy,
compares it with an incumbent, marks the branch, and uncomputes. BBHT/D\"urr--H{\o}yer minimum finding
~\cite{boyer1998tight,durr1996minimum} then calls this oracle
\(O(\sqrt{N_q})\) times. No candidate energy is evaluated or stored
individually; each oracle call acts on all parameter branches in
superposition. Here ``deformation'' refers to the parametric deformation of
the Hamiltonian family, of which nuclear shape deformation is one instance.

The reduction concerns the number of many-body solves, and its relevance
depends on two quantities that we make explicit. The first is the constant:
for the ideal search layer we measure \(7.4\sqrt{N_q}\) queries for 99\%
success, so the query count falls below \(N_q\) only for
\(N_q\gtrsim55\). The second is the overhead \(\xi\) of one coherent
oracle call relative to one pointwise solve. With both included, QDMin
needs fewer solve-equivalents than a pointwise quantum scan for
\(N_q\gtrsim220\xi^2\) (Sec.~\ref{sec:pointwise-scan}). One-dimensional
landscapes with a few dozen points therefore gain nothing. Multidimensional
order-parameter spaces do: five quadrupole channels on a 16-point grid per
channel give \(N_q\approx10^6\). Exploitable classical structure, such as
symmetry reduction, smoothness, and continuation, reduces \(N_q\) for both
approaches and is part of the comparison
~\cite{walesdoye1997basinhopping,jones1998ego}.

Quantum-computing studies in nuclear physics have primarily targeted states,
energies, or observables of a specified Hamiltonian, from few-body
demonstrations~\cite{dumitrescu2018nucleus} to variational ground-state
constructions~\cite{stetcu2022variational}, ordered coupled-cluster
circuits~\cite{kiss2022ordereducc}, and adaptive ansatzes
~\cite{romero2022adapt}. Projection-based preparation and structured pairing
encodings address parts of the state-access cost
~\cite{stetcu2023projection,liu2025pairingblockencoding}, while adiabatic and
annealing proposals expose the importance of path gaps
~\cite{costa2025annealing,costa2026quasiparticle}. Pointwise deformation
calculations on quantum computers have used separate variational solves at
fixed deformation values~\cite{abhishek2026crankedzr}; that is the pointwise
scan QDMin is compared with.

Quantum global optimization includes Grover adaptive search and quantum basin
hopping~\cite{baritompa2005grover,bulger2005basinhopping}, simulated
searches over classically evaluated mesh values~\cite{zhu2009simulated}, and
objective-loading schemes followed by variational search
~\cite{soltaninia2025qgmf}. Energy-profile learning and molecular-geometry
optimization use classically updated parameters
~\cite{cerveralierta2021metavqe,delgado2021geometry,
yuan2021hamiltonianwavefunction}. The closest many-body precedents are
molecular. Kassal and Aspuru-Guzik combine local geometry optimization with
minimum search over basin representatives~\cite{kassal2009geometry}, and
Kosugi, Nishi, and Matsushita encode geometries coherently and concentrate
the geometry histogram by probabilistic imaginary-time evolution
~\cite{kosugi2023geometry}; electron--nuclear quantum dynamics offers a
further route~\cite{hirai2022electronsnuclei}. Coherent parameter registers
and quantum global optimization are therefore established. QDMin contributes
an on-demand ground-energy threshold oracle for incumbent-based minimum
finding over a parametrized interacting family, together with the cost model
that decides when it pays.

Our contributions are:
\begin{itemize}
\item \emph{A problem class.} Static-field families, in which the
parametrized minimum is a rigorous variational bound that the parent ground
state does not reveal (Sec.~\ref{sec:problem}).
\item \emph{A coherent oracle that tolerates imperfect preparation.}
Imperfect ground states can make the threshold test miss a branch but not
accept a wrong one; fixed-point amplification repairs the misses at a cost
that grows only logarithmically with the target error
(Sec.~\ref{sec:robust-predicate}).
\item \emph{An honest cost model.} The measured search constant gives the
grid size at which QDMin needs fewer solves than a pointwise quantum scan
(Sec.~\ref{sec:resources}).
\item \emph{Realistic tests.} Shell-model (USDB) landscapes of
\(^{20}\)Ne, \(^{24}\)Mg, and \(^{28}\)Si with up to \(10^6\) points, and a
Hubbard ring, show where the search pays and where it does not
(Secs.~\ref{sec:sd-shell} and~\ref{sec:hubbard}).
\item \emph{Circuit-level validation.} The complete pipeline runs with
circuit-generated energies on a schematic model in one and two coordinates
(Sec.~\ref{sec:validation}).
\end{itemize}

\begin{figure*}[tbp]
  \centering
  \includegraphics[width=0.96\textwidth]{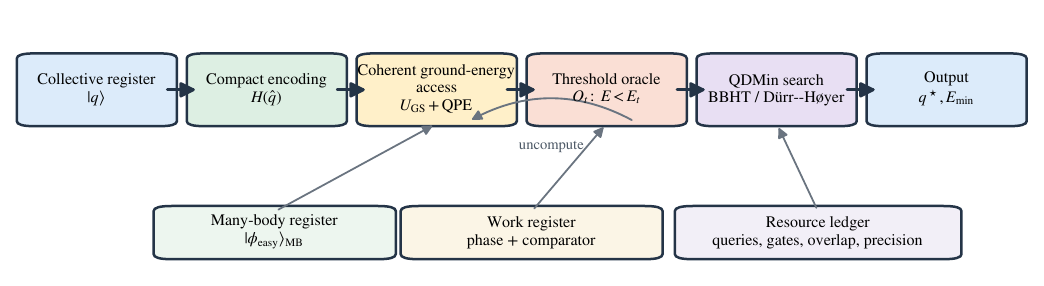}
  \caption{QDMin architecture. The main path keeps the parameter coordinates
  coherent while compact \(H(\hat{\bm q})\), ground-state access, quantum phase
  estimation (QPE),
  reversible comparison, and uncomputation generate a below-incumbent phase
  predicate. BBHT/D\"urr--H{\o}yer minimum finding
  ~\cite{boyer1998tight,durr1996minimum} then updates the incumbent
  and returns \((\bm q^\star,E_{\min})\). The many-body, phase, and comparator
  register costs are accounted for separately from candidate-query counts.}
  \label{fig:algorithm-schematic}
\end{figure*}

\section{Problem formulation}
\label{sec:problem}

\subsection{Parametrized minimum}

Let \(\bm q=(q_1,\ldots,q_d)\) be a small set of parameters and consider the
family
\begin{equation}
\begin{gathered}
H(\bm q)
=
H_0+\sum_{\mu=1}^{R} f_\mu(\bm q) H_\mu ,
\\
E_0(\bm q)=\min\operatorname{spec}H(\bm q),
\end{gathered}
\label{eq:general-family}
\end{equation}
with fixed Hermitian components \(H_\mu\) and efficiently computable
coefficient functions \(f_\mu\). QDMin seeks
\begin{equation}
\bm q^\star
=
\underset{\bm q\in\mathcal G}{\arg\min}\,
E_0(\bm q)
\end{equation}
over a finite grid \(\mathcal G\) of \(N_q\) candidates. The problem is to
identify which member of an interacting family has the lowest ground-state
energy, not to find the ground state of one fixed Hamiltonian.

\subsection{Static-field families}

The families studied here arise from a parent Hamiltonian with attractive
quadratic channels,
\begin{equation}
H=H_0-\sum_{a=1}^{d}\frac{\chi_a}{2}A_a^2,
\qquad \chi_a>0,\quad A_a=A_a^\dagger .
\label{eq:parent}
\end{equation}
Replacing each \(A_a^2\) by its static decoupling gives
\begin{equation}
H(\bm q)
=
H_0-\sum_a\chi_a q_a A_a+\sum_a\frac{\chi_a}{2}q_a^2 I ,
\label{eq:static-field-family}
\end{equation}
which has the form of Eq.~\eqref{eq:general-family} with \(R=2d\) and
coefficients of degree at most two. The field \(\bm q\) is the
time-independent restriction of the Hubbard--Stratonovich auxiliary field
~\cite{stratonovich1957method,hubbard1959partition}. Static fields are
integrated over at finite temperature in the static-path approximation
~\cite{arve1988staticpath}, and time-dependent fields are sampled in
shell-model Monte Carlo~\cite{koonin1997smmc}. Here a single static field is
optimized at zero temperature, for which the following bound holds.

\emph{Variational bound.} For every real \(\bm q\),
\begin{equation}
H(\bm q)-H=\sum_a\frac{\chi_a}{2}\left(A_a-q_aI\right)^2\succeq0 .
\label{eq:operator-bound}
\end{equation}
The \(A_a\) need not commute with one another or with \(H_0\). Hence
\(E_0(H)\leq E_0(\bm q)\) for all \(\bm q\), and
\(\min_{\bm q}E_0(\bm q)\) is the best upper bound to \(E_0(H)\) obtainable
with a static field in the decoupled channels. All other correlations,
including pairing, are retained exactly.

\emph{Self-consistency.} For a nondegenerate ground-state branch, the
Hellmann--Feynman relation~\cite{feynman1939forces} gives
\begin{equation}
\frac{\partial E_0}{\partial q_a}=\chi_a\left(q_a-\langle A_a\rangle_{\bm q}\right),
\label{eq:general-hellmann-feynman}
\end{equation}
so an interior minimum satisfies
\begin{equation}
q_a^\star=\langle A_a\rangle_{\bm q^\star}.
\label{eq:self-consistency}
\end{equation}
The optimal field is the self-consistent order parameter of the decoupled
channels.

\emph{Why the family must be searched.} The ground state of the parent
Hamiltonian does not provide \(\bm q^\star\). If \(H\) is invariant under a
symmetry that transforms the \(A_a\) nontrivially, its nondegenerate ground
state has \(\langle A_a\rangle=0\). For a rotationally invariant nuclear
Hamiltonian with quadrupole channels \(A_a\) built from \(Q_{2\mu}\), the
\(J=0\) ground state carries no intrinsic quadrupole moment, whereas
\(\bm q^\star\) is the intrinsic deformation. The same holds for magnetic or
charge order parameters of finite lattice models. Locating \(\bm q^\star\)
therefore requires the parametrized search; it cannot be replaced by
ground-state minimization of \(H\).

\emph{Linear field couplings.} Any family
\(H_0-\kappa qA+\tfrac{C}{2}q^2I\) with \(\kappa\neq0\) and \(C>0\) is of the
form~\eqref{eq:static-field-family}, with \(\chi=\kappa^2/C\) and field
coordinate \((C/\kappa)q\). Collective-field Hamiltonians with a quadratic restoring
term are therefore static-field decouplings of
\(H_0-(\kappa^2/2C)A^2\).

\subsection{Nuclear example}

The circuit-level demonstrator uses two fermions in four spin-orbitals
arranged as two time-reversed (Kramers) pairs~\cite{kramers1930paramagnetic},
\((0,1)\) and \((2,3)\). In the occupation basis, with a
Jordan--Wigner mapping~\cite{jordanwigner1928mapping} for the circuit
representation,
\begin{equation}
H(q)=H_{\rm sp}-G P^\dagger P-\kappa q Q
  +\frac{C}{2}q^2 I .
\label{eq:model-hamiltonian}
\end{equation}
Pairing combined with a separable quadrupole interaction is an established
schematic framework for correlated nuclear collective motion
~\cite{barangerkumar1968pairingquadrupole}. Here
\(P^\dagger=c_0^\dagger c_1^\dagger+c_2^\dagger c_3^\dagger\) creates a
correlated pair in either time-reversed orbital pair, and the diagonal
one-body operator \(Q=\sum_p m_p n_p\), with schematic moments
\(m=(-1,-1,+1,+1)\), provides a minimal quadrupole-like polarization. The
coordinate \(q\) plays the role of the quadrupole deformation \(\beta_2\).

By the linear-coupling identity, Eq.~\eqref{eq:model-hamiltonian} is the
static-field family of the pairing-plus-quadrupole Hamiltonian
\(H_{\rm sp}-GP^\dagger P-(\kappa^2/2C)Q^2\). For the parameters used below
(\(G=0.7\), \(\kappa=0.8\), \(C=2.1\)), the continuous minimum is
\(E_0=-2.6098\) at \(q=-0.731\), above the parent ground energy \(-2.6988\),
and the smallest eigenvalue of \(H(q)-H\) is zero, so the bound is tight in
the operator sense. In the units of Eq.~\eqref{eq:model-hamiltonian}, the
Hellmann--Feynman relation reads
\begin{equation}
\frac{dE_0}{dq}=-\kappa\langle Q\rangle_q+Cq,
\label{eq:hellmann-feynman-balance}
\end{equation}
and the self-consistency condition~\eqref{eq:self-consistency} becomes
\begin{equation}
Cq^\star=\kappa\langle Q\rangle_{q^\star}.
\label{eq:stationary-field-balance}
\end{equation}
The deformation content of \(q\) is thus precise: it is the
self-consistent quadrupole field of a pairing-plus-quadrupole Hamiltonian
whose pairing correlations are treated exactly.

The four spin-orbitals map to a four-qubit Fock space of dimension
\(2^4=16\). All implemented dynamics conserve particle number, so the
physical evolution is restricted to the \(N=2\) subspace of dimension
\(\binom{4}{2}=6\). Energies are in arbitrary consistent units. The model is
deliberately small: it is the largest instance for which the complete
coherent oracle, including phase estimation and uncomputation, can be
simulated at the circuit level. Realistic families, a shell-model
quadrupole field and a Hubbard ring with competing orders, follow in
Secs.~\ref{sec:sd-shell} and~\ref{sec:hubbard}.

\section{Compact coherent parameter encoding}

The goal is to let a quantum register represent all candidate values of the
collective coordinate while avoiding a separate controlled Hamiltonian for
every grid point. For \(N_q=2^{n_q}\) uniformly spaced values
\[
q_k=q_{\min}+k\Delta q,\qquad k=0,\ldots,N_q-1,
\]
we encode the coordinate itself as the diagonal operator
\begin{equation}
\hat q=q_{\min}I+\Delta q\sum_{r=0}^{n_q-1}
2^r\frac{I-Z_r}{2}.
\end{equation}
On the computational-basis state representing the integer \(k\),
\(\hat q\) therefore has eigenvalue \(q_k\).

The model introduced above depends quadratically on the collective field,
\begin{equation}
H(q)=H_0+qH_1+q^2H_2,
\end{equation}
with
\[
H_0=H_{\rm sp}-GP^\dagger P,\quad
H_1=-\kappa Q,\quad
H_2=\tfrac{C}{2}I.
\]
Replacing the scalar \(q\) by the register operator \(\hat q\) gives the
coherent Hamiltonian
\begin{equation}
\mathcal H
=
I\otimes H_0
+\hat q\otimes H_1
+\hat q^2\otimes H_2.
\end{equation}
Each coordinate basis state therefore selects the corresponding member
\(H(q_k)\) of the many-body Hamiltonian family, without introducing an
explicit projector \(\lvert k\rangle\langle k\rvert\) for every candidate.

For the 16-point validation grid,
\[
q_k=-1.2+0.16k,\qquad k=0,\ldots,15,
\]
the grid is centered around zero. In this case the constant contribution in
\(\hat q\) cancels, leaving
\[
\hat q=-0.08
\left(Z_0+2Z_1+4Z_2+8Z_3\right).
\]
Thus \(\hat q\) contains four Pauli-\(Z\) monomials. Squaring this operator
produces one identity term and the six distinct pair products
\(Z_rZ_s\), giving seven terms in \(\hat q^2\).

The four-point dynamic grid,
\[
(-0.88,-0.72,-0.56,-0.40),
\]
is not centered around zero. Its two-qubit coordinate operator therefore
contains an identity term in addition to \(Z_0\) and \(Z_1\), giving three
terms in \(\hat q\), while \(\hat q^2\) contains four. Together with the
physical Pauli decompositions of \(H_0\), \(H_1\), and \(H_2\), containing
15, 4, and 1 terms, respectively, the complete coherent Hamiltonian contains
26 distinct Pauli strings after identical products are combined. The dense
reconstruction residual is \(8.88\times10^{-16}\).

\begin{figure}[tbp]
  \centering
  \includegraphics[width=\columnwidth]{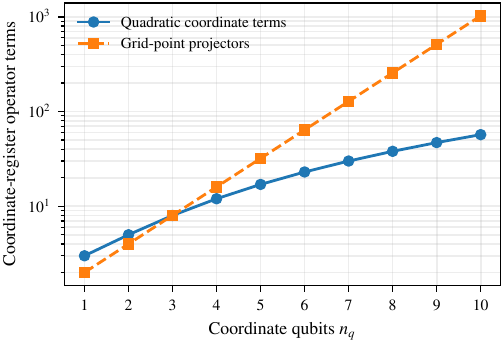}
  \caption{Coordinate-register representation cost for a centered quadratic
Hamiltonian family compared with explicit grid-point control. The polynomial
construction requires \(2+n_q+n_q(n_q-1)/2\) constant, linear, and quadratic
coordinate-monomial contributions, whereas
\(\mathcal H_{\rm proj}=\sum_k |k\rangle\langle k|\otimes H(q_k)\)
contains \(2^{n_q}\) explicitly enumerated coordinate projectors. At
\(n_q=10\), these counts are 57 and 1024, respectively. The comparison
concerns only the representation of the coordinate dependence and excludes
the physical Pauli decomposition, circuit-synthesis cost, and energy-oracle
cost.}
  \label{fig:encoding-scaling}
\end{figure}

Figure~\ref{fig:encoding-scaling} extends this coordinate-operator count
through ten coordinate qubits. For a centered quadratic family, the constant,
linear, and quadratic pieces contribute
\[
1,\qquad n_q,\qquad
1+\frac{n_q(n_q-1)}{2}
\]
coordinate Pauli monomials, respectively, giving
\[
2+n_q+\frac{n_q(n_q-1)}{2}
\]
terms before combining identical identity contributions across the different
Hamiltonian components.

For comparison, a direct grid-point construction would write the coherent
Hamiltonian as
\begin{equation}
\mathcal H_{\rm proj}
=
\sum_{k=0}^{N_q-1}
|k\rangle\langle k|\otimes H(q_k),
\qquad
N_q=2^{n_q}.
\end{equation}
This representation contains one explicitly specified coordinate projector,
and hence one separately conditioned Hamiltonian block, for every grid point.
The number of such coordinate controls therefore grows as \(2^{n_q}\).
For example, at \(n_q=10\), the quadratic construction contains
\(2+10+45=57\) coordinate monomial contributions, whereas the direct
grid-point construction contains \(2^{10}=1024\) projectors.

Figure~\ref{fig:encoding-scaling} compares the representation cost of the
coordinate dependence. The physical-system Pauli decomposition, controlled
operations, and ground-energy oracle contribute separately to the total cost.

For two schematic field coordinates we use independent registers, with
\(q_2\) as the low subregister and \(q_3\) as the high subregister. Thus
\(k=i_2+N_2i_3\), and Euclidean division reversibly recovers \((i_2,i_3)\).
The encoded family is
\begin{equation}
\mathcal H=\sum_{p_2,p_3}\hat q_2^{p_2}
\hat q_3^{p_3}\otimes H_{p_2p_3}.
\end{equation}
For the five polynomial components used below and two qubits per coordinate,
the combined coordinate factors have seven distinct Pauli-\(Z\) supports,
versus 16 individual grid-point projectors. This count again excludes the
physical-operator decomposition.

Binary encoding compresses \(N_q=2^{n_q}\) candidate labels into \(n_q\)
qubits. For a fixed-degree polynomial family, the coordinate Pauli expansion
also grows polynomially in \(n_q\). Computational advantage depends on the
cost of implementing and querying the resulting Hamiltonian.

\section{Coherent ground-energy and threshold oracle}

\subsection{Deformation-coherent preparation}

The threshold oracle requires coherent ground-state access on each
collective-coordinate branch, followed by quantum phase estimation
(QPE)~\cite{cleve1998algorithms,abrams1999eigenvalues}.
For a general input state
\begin{equation}
|\phi(q)\rangle
=
\sum_j c_j(q)|\psi_j(q)\rangle ,
\end{equation}
QPE coherently associates each eigencomponent with its phase. A
measurement samples the corresponding outcomes with weights
\(|c_j(q)|^2\). Ground-state preparation or filtering concentrates this weight
on the lowest-energy component
~\cite{poulinwocjan2009ground,lintong2020filter,lintong2020ground}. We denote
this operation by \(U_{\rm GS}\); the present demonstrator realizes it with
adiabatic state preparation (ASP)~\cite{jansen2007adiabatic}.

For the four-orbital model, the initial state \(|\phi_{\rm easy}\rangle\)
is the analytically known \(N=2\) ground determinant of the diagonal
single-particle Hamiltonian \(H_{\rm sp}\), with the two lower single-particle
orbitals occupied. ASP uses this state on every coordinate branch and turns on
the interactions while the coordinate register remains coherent. If the
schedule is slow enough relative to both
the branch gaps and the variation of the Hamiltonian along the path, each
branch follows the corresponding low-energy eigenstate of its own Hamiltonian.
A superposition of collective coordinates is then transformed approximately as
\begin{equation}
\sum_q a_q |q\rangle |\phi_{\rm easy}\rangle
\;\longrightarrow\;
\sum_q a_q |q\rangle |\widetilde{\psi}_0(q)\rangle ,
\end{equation}
without measuring or classically selecting the coordinate \(q\).

The preparation path is
\begin{equation}
H_{\rm prep}(s,\hat q)
=
H_{\rm sp}
+
s\left[
-GP^\dagger P-\kappa \hat q Q
\right]
\label{eq:prep-path}
\end{equation}
with \(0\leq s\leq1\).
At \(s=0\), every branch therefore begins from the same simple determinant.
As \(s\) increases, the pairing interaction and the
coordinate-dependent collective field are turned on simultaneously, so the
many-body state on each branch evolves toward the corresponding interacting
low-energy state.

The restoring contribution \(C\hat q^2 I/2\) is omitted from
Eq.~\eqref{eq:prep-path} because it is proportional to the identity in the
many-body space. It shifts the energy, and therefore accumulates a
coordinate-dependent phase, but it does not change the many-body
eigenvectors that ASP is intended to prepare.

The continuous path is implemented using a linear schedule in \(s\). At each
midpoint, \(H_{\rm prep}(s,\hat q)\) is assembled directly from the compact
Pauli representation and evolved with a second-order Suzuki product
formula~\cite{trotter1959product,suzuki1990fractal}.
Because \(\hat q\) remains an operator on the coordinate register, the same
circuit acts simultaneously on an arbitrary superposition of coordinate
values. Oracle uncomputation~\cite{bennett1973reversibility} later applies
the exact inverse of this implemented preparation sequence.

For the calculations reported below, the total adiabatic evolution time is
\(T=4\) in inverse model-energy units. The path is divided into eight
midpoint slices, with one second-order Suzuki repetition per slice. Finite
evolution time and product-formula error limit the fidelity of
\(|\widetilde{\psi}_0(q)\rangle\) to the exact ground state
\(|\psi_0(q)\rangle\). Adiabatic accuracy depends on the spectral
gap along the entire preparation path
~\cite{jansen2007adiabatic}. Related nuclear annealing studies provide
fixed-Hamiltonian precedents for this sensitivity
~\cite{costa2025annealing,costa2026quasiparticle}. In QDMin, the collective
coordinate operator \(\hat q\) controls the adiabatic path coherently.

The quality of this ground-state access is evaluated only after the QDMin
search result has been frozen. Exact eigenvectors are then used to compute
the branch fidelities
\begin{equation}
F_q
=
\left|
\langle
\psi_0(q)
|
\widetilde{\psi}_0(q)
\rangle
\right|^2
\end{equation}
and to inspect the minimum instantaneous spectral gap along each preparation
path. These exact-state comparisons provide retrospective validation.

\subsection{Certified spectral interval}

QPE represents energy through a phase, and phases are periodic. The energy
range used to define the phase map must therefore contain every eigenvalue
that can be reached during the calculation; otherwise two physically
different energies could wrap onto the same phase and the subsequent energy
comparison would become ambiguous. We construct the interval directly from
the analytic Hamiltonian coefficients, keeping its evaluation independent of
candidate eigensolutions. Writing the
coordinate-dependent Hamiltonian in polynomial form,
\begin{equation}
H(q)=\sum_p q^p H_p,
\end{equation}
we restrict each component \(H_p\) to the particle-number-conserving
\(N=2\) subspace. This restriction is sufficient because the initial
determinant contains exactly two particles and both the preparation path and
the target Hamiltonian conserve particle number. The physical evolution
therefore remains inside this invariant six-dimensional subspace even though
the circuit itself acts on the full four-qubit Fock space.

To bound the spectrum throughout the entire continuous coordinate interval
\([q_{\min},q_{\max}]\), we apply Gershgorin's theorem
~\cite{gershgorin1931bounds,hornjohnson2012matrix}. For matrix row \(i\),
the diagonal entry is a
real polynomial in \(q\), whose minimum and maximum over the interval can be
obtained analytically. The total magnitude of the off-diagonal entries is
bounded by
\begin{equation}
R_i(q)
\leq
\overline R_i
=
\sum_p
\max_{q\in[q_{\min},q_{\max}]}
|q|^p
\sum_{j\ne i}|(H_p)_{ij}|.
\end{equation}
If \(d_i^-\) and \(d_i^+\) denote the minimum and maximum, respectively, of
the diagonal polynomial for row \(i\), then
\begin{equation}
E_L
=
\min_i\left(d_i^- -\overline R_i\right),
\qquad
E_U
=
\max_i\left(d_i^+ +\overline R_i\right)
\label{eq:gershgorin-bound}
\end{equation}
bounds every eigenvalue of the \(N=2\) Hamiltonian for every
\(q\in[q_{\min},q_{\max}]\).

The bounds \(E_L\) and \(E_U\) use the known matrix or Pauli coefficients.
A regression test verifies their independence from candidate diagonalization
by constructing the interval with all repository eigensolver entry points
disabled.

For the four-point search interval, Eq.~\eqref{eq:gershgorin-bound} gives
\[
[E_L,E_U]
=
[-3.20952381,\,3.42112000].
\]
The corresponding affine operator
\begin{equation}
A
=
\frac{\mathcal H-E_L I}
{(E_U-E_L)/(1-\eta)},
\qquad
\eta=0.125,
\end{equation}
maps the certified lower bound to phase zero and the certified upper bound to
phase \(1-\eta\). The parameter \(\eta\) therefore leaves a guard interval
before the periodic phase boundary at unity. With nine phase qubits, the
resulting energy-bin width is \(0.0148005\).

After freezing the QDMin result, we diagonalize the candidate Hamiltonians
to check the certificate. All postvalidated \(N=2\) eigenphases lie between
\(0.07916\) and \(0.79481\), well inside the certified interval. The
coefficient-based Gershgorin bound thus supplies the phase map, while the
post-search spectra validate it.

\subsection{Structured QPE and approximate threshold comparison}

The prepared state on a coordinate branch generally contains several
eigencomponents,
\[
|\widetilde{\psi}_0(q)\rangle
=\sum_j c_j(q)|\psi_j(q)\rangle,
\qquad |c_0(q)|^2=F_q.
\]
For exact Hamiltonian evolution, finite-precision QPE gives
\begin{equation}
\begin{aligned}
&\sum_q a_q |q\rangle|\widetilde{\psi}_0(q)\rangle
|0\rangle_{\rm phase}
\\
&\quad\longmapsto
\sum_{q,j}a_qc_j(q)|q\rangle|\psi_j(q)\rangle
|\Phi_m(\phi_j(q))\rangle_{\rm phase}.
\end{aligned}
\label{eq:qpe-eigencomponents}
\end{equation}
Here \(\phi_j(q)\) is the affine phase associated with \(E_j(q)\), and
\[
|\Phi_m(\phi_j(q))\rangle
=\sum_{k=0}^{2^m-1}b_k(\phi_j(q))|k\rangle
\]
is the normalized \(m\)-qubit QPE output. It reduces to one basis state only
when the eigenphase is exactly representable. Otherwise, it has probability
concentrated near the corresponding phase bin, with tails at other bins.
Different eigencomponents generally remain entangled with different phase
estimates. An exact eigenstate input gives a product of the unchanged
many-body state and its phase-register state.

Measured incumbent updates use the modal bin from a QPE shot batch. The
coherent threshold oracle acts on the entire phase-register state; it does
not select that modal bin or project onto the ground component.
Product-formula simulation adds error to the exact-evolution description in
Eq.~\eqref{eq:qpe-eigencomponents}.

The formal QPE construction follows
~\cite{cleve1998algorithms,abrams1999eigenvalues}. Recent analyses clarify
the coherent use of phase estimates in downstream quantum circuits
~\cite{lintong2022heisenberg,patel2026taperedqpe}, while randomized evolution
provides a different resource tradeoff~\cite{gunther2026randomizedqpe}. Here
and below, ``structured'' means that each controlled binary power required
by QPE is synthesized from the compact Pauli decomposition of the coherent
Hamiltonian using a second-order Suzuki product formula
~\cite{trotter1959product,suzuki1990fractal,childs2021trotter}. Dense exact
evolution provides a retrospective validation reference. Quantum signal
processing and qubitization offer alternative simulation constructions
~\cite{lowchuang2017qsp,lowchuang2019qubitization}.

Once QPE has written the energy estimate, QDMin asks whether the candidate
energy is lower than the current incumbent \(E_t\).
Because the affine phase map is monotonic in energy, all phase bins satisfying
\(\widetilde E<E_t\) form an initial interval
\([0,k_t)\) of the binary phase register. We decompose this interval into at
most \(m\) aligned binary prefixes, where \(m\) is the number of phase
qubits, and apply a phase flip to each prefix. This implements the comparison
without introducing flag or carry ancillas and without constructing a dense
diagonal operator.

After marking, the circuit applies inverse QPE followed by inverse
preparation. For a fixed coordinate \(q\), let \(W_q\) denote the implemented
preparation followed by QPE, and let \(S_t\) flip the marked phase bins. The
work-register operation is \(W_q^\dagger S_tW_q\), whose action can be
written as
\begin{equation}
\begin{aligned}
&W_q^\dagger S_tW_q|0\rangle_{\rm work}
\\
&\qquad=z_t(q)|0\rangle_{\rm work}+|g_t(q)\rangle,
\end{aligned}
\label{eq:threshold-work-residual}
\end{equation}
where \(\langle0|g_t(q)\rangle=0\) and
\(|z_t(q)|^2+\|g_t(q)\|^2=1\).
If \(p_t(q)\) is the marked probability weight after \(W_q\), then
\(z_t(q)=1-2p_t(q)\).

A clean ground-energy predicate requires all populated components on a
coordinate branch to receive the correct common sign. For an exact ground
state and an error-free energy comparison, the ideal action is
\begin{equation}
\begin{aligned}
|q\rangle|0\rangle_{\rm work}
\longmapsto
(-1)^{[E_0(q)<E_t]}|q\rangle|0\rangle_{\rm work}.
\end{aligned}
\label{eq:ideal-threshold-action}
\end{equation}
Preparation admixtures can place eigencomponents on opposite sides of the
threshold, and finite-resolution QPE can place bins from one eigencomponent
on both sides. Selective phase marking then changes the state entering the
inverse operations, so exact inversion of the implemented circuits does not
by itself guarantee clean work registers. Product-formula error further
affects the phase estimates and their comparison.

The reported search simulations retain the full work state and measure its
clean probability explicitly. Cleanup and predicate correctness are distinct:
a branch can return cleanly with the wrong sign. Ground-state fidelity alone
also does not specify the threshold error, which depends on the incumbent
and the phase distributions. These diagnostics characterize the implemented
finite circuits; they do not supply a worst-case oracle-error bound or a
bounded-error minimum-search guarantee.

\subsection{Overlap-robust threshold predicate}
\label{sec:robust-predicate}

The threshold test above inherits every imperfection of the prepared state.
If a branch lies below the threshold but its prepared state has ground-state
fidelity \(F_q\), only that fraction of the state is marked, and the branch
returns with amplitude \(z_t(q)\approx1-2F_q\) instead of \(-1\). At
\(F_q=0.75\) the branch is flipped only halfway. The search then works with a
blurred oracle, and the error grows with the number of calls.

The test has a useful asymmetry. Every excited state of \(H(q)\) lies above
its ground state, so an imperfect preparation can only add weight at higher
energies. A branch whose ground energy is above the threshold therefore
cannot be marked by its excited admixtures; it can be marked only through the
small tails of the phase estimate. A branch whose ground energy is below the
threshold is always marked with at least the weight of its ground component.
Imperfect preparation can make the test miss a branch, but it cannot make it
accept a wrong one. If \(\ell(\varepsilon)\) denotes the largest weight that
the phase estimate of an eigencomponent places on the wrong side of a
threshold at distance \(\varepsilon\) or more, the marked weight \(p_t(q)\)
obeys
\begin{equation}
\begin{aligned}
E_0(q)&\geq E_t+\varepsilon &&\Longrightarrow\quad p_t(q)\leq\ell(\varepsilon),\\
E_0(q)&\leq E_t-\varepsilon &&\Longrightarrow\quad p_t(q)\geq F_q\,[1-\ell(\varepsilon)].
\end{aligned}
\label{eq:one-sided}
\end{equation}
Branches within \(\varepsilon\) of the threshold cannot be decided at this
precision, as in any comparison of finite accuracy.

A miss can be repaired by amplifying the marked part of the state before the
sign is applied. Ordinary amplitude amplification needs to know how large
that part is, and it is different on every branch. Fixed-point amplitude
amplification~\cite{yoder2014fixedpoint} does not: with \(L\) uses of the
preparation it raises every marked weight above a chosen floor \(w\) to
nearly one, and leaves a weight of zero at zero. For marked weight \(p\) it
produces
\begin{equation}
P_L(p)=1-\delta^2T_L^2\!\left(T_{1/L}(1/\delta)\sqrt{1-p}\right),
\label{eq:fixed-point-success}
\end{equation}
where \(T_L\) is a Chebyshev polynomial. This is at least \(1-\delta^2\)
for every \(p\geq w\) once \(L\geq\log(2/\delta)/\sqrt{w}\). Applying the
sign between the amplified preparation and its inverse gives a branch
amplitude \(1-2P_L(p_t(q))\): within \(2\delta^2\) of the ideal \(-1\) on
branches clearly below the threshold, and close to \(+1\) on branches clearly
above it. The price is \(2L\) applications of preparation and phase
estimation per call instead of two. Because \(L\) grows only as
\(\log(1/\delta)/\sqrt{F_{\min}}\), the sequence stays short: a guaranteed
fidelity of \(0.5\) and a target error of \(\delta^2=0.01\) need \(L=5\).

In circuit form, with \(A_q\) the preparation followed by phase estimation,
the amplified state is
\(V_q=G(\alpha_l,\beta_l)\cdots G(\alpha_1,\beta_1)A_q\) with \(L=2l+1\) and
\(G(\alpha,\beta)=-A_qS_0(\alpha)A_q^\dagger S_t(\beta)\). Here
\(S_0(\alpha)=I-(1-e^{-i\alpha})|0\rangle\langle0|\) acts on the work
registers, \(S_t(\beta)\) multiplies the marked phase bins by \(e^{i\beta}\)
using the same prefix construction as the comparator, and the angles are
those of Ref.~\cite{yoder2014fixedpoint}. The oracle is
\(V_q^\dagger S_t(\pi)V_q\), uncomputed by the exact circuit inverse. The
coordinate register is never touched, so all branches are amplified in
parallel.

There is one catch. Amplification cannot tell a small genuine mark from a
small spurious one, so the tails of the phase estimate are amplified too,
roughly by a factor \(L^2\). The robust test is useful only if
\(\ell(\varepsilon)\ll L^{-2}\). Textbook phase estimation, which starts the
phase register in a uniform superposition, has heavy tails: a component two
bins away from a threshold can still leak several per cent of its weight
across it. Starting the phase register instead in a smoothly tapered state
\(\sum_kw_k|k\rangle\)~\cite{rendon2022tapering,patel2026taperedqpe}
suppresses these tails by orders of magnitude, at the cost of a slightly
wider central peak. We use a Kaiser window, prepared on the phase register by
a binary tree of controlled \(R_y\) rotations.

\subsection{General parameterized many-body threshold oracle}

The general logical target is a clean below-incumbent predicate for a
parameterized interacting many-body family, subject to the preparation and
comparison-accuracy requirements above.

Let
\(\bm\lambda=(\lambda_1,\ldots,\lambda_d)\) denote a set of external or
collective parameters, and consider a Hamiltonian family
\begin{equation}
H(\bm\lambda)
=
\sum_{\mu=0}^{R}
f_\mu(\bm\lambda) H_\mu .
\label{eq:general-parameter-family}
\end{equation}
Here the \(H_\mu\) are fixed many-body operator components, while the
functions \(f_\mu(\bm\lambda)\) specify how their strengths depend on the
parameters. The integer \(R+1\) counts the operator components in the
chosen parameterization.

To search over the parameters coherently, each \(\lambda_i\) is represented
by a quantum register and promoted to a commuting diagonal operator
\(\hat\lambda_i\). The corresponding coherent Hamiltonian is then
\begin{equation}
\mathcal H(\hat{\bm\lambda})
=
\sum_{\mu=0}^{R}
f_\mu(\hat{\bm\lambda})\otimes H_\mu .
\label{eq:general-coherent-hamiltonian}
\end{equation}
On a computational-basis state
\(|\bm\lambda\rangle\), this operator acts with the many-body Hamiltonian
\(H(\bm\lambda)\) associated with that parameter value. The usefulness of
this construction depends on structure: the functions
\(f_\mu(\hat{\bm\lambda})\) and the physical operators \(H_\mu\) must admit
a sufficiently compact implementation.

The next requirement is coherent access to the corresponding low-energy
state. We denote the abstract ground-state-access operation by
\begin{equation}
U_{\rm GS}:
|\bm\lambda\rangle|0\rangle
\longmapsto
|\bm\lambda\rangle
|\widetilde{\psi}_0(\bm\lambda)\rangle .
\label{eq:general-ground-state-access}
\end{equation}
The \(U_{\rm GS}\) interface accommodates different state-preparation or
filtering strategies; adiabatic preparation is the realization used here.

In the ideal limit of exact ground-state access and eigenphases exactly
representable on the phase grid, the spectral interface is
\[
\begin{aligned}
&|\bm\lambda\rangle|\psi_0(\bm\lambda)\rangle|0\rangle_{\rm phase}
\\
&\quad\longmapsto
|\bm\lambda\rangle|\psi_0(\bm\lambda)\rangle|k_0(\bm\lambda)\rangle_{\rm phase},
\end{aligned}
\]
where \(k_0(\bm\lambda)\) is the binary phase label decoding to
\(E_0(\bm\lambda)\). Reversible comparison and uncomputation then give
\begin{equation}
O_t^{\rm ideal}|\bm\lambda\rangle
=
(-1)^{[E_0(\bm\lambda)<E_t]}|\bm\lambda\rangle.
\label{eq:general-threshold-oracle}
\end{equation}
Approximate preparation and finite phase resolution instead follow
Eq.~\eqref{eq:qpe-eigencomponents} and can leave the work residual in
Eq.~\eqref{eq:threshold-work-residual}. Using the implemented operation as a
coordinate-only phase oracle therefore requires control of both the
comparison error and the residual work.

With the ideal predicate, BBHT/D\"urr--H{\o}yer minimum finding acts only
on the parameter register, independently of whether the Hamiltonian describes
nuclear structure, electronic structure, or another interacting many-body
problem. The finite demonstrations test this interface with an approximate
oracle, whose errors require separate analysis to establish a search guarantee.

What changes from one physical problem to another is the cost of realizing
those ingredients: the parameter dependence, many-body operator structure,
ground-state access, Hamiltonian simulation, spectral normalization, gaps,
and required energy precision are all model dependent. Replacing the present
four-orbital benchmark by a more realistic Hamiltonian preserves the logical
threshold-search structure while changing its implementation cost.

As one nuclear example, consider a shell-model
Hamiltonian~\cite{caurier2005shellmodel}
\begin{equation}
H_{\rm SM}
=
\sum_{ab} t_{ab} a_a^\dagger a_b
+
\frac14
\sum_{abcd}
V_{abcd}
a_a^\dagger a_b^\dagger a_d a_c .
\end{equation}
A collective quadrupole field can define the parameterized family
\begin{equation}
H_{\rm SM}(q)
=
H_{\rm SM}
-\kappa q Q_{20}
+\frac{C}{2}q^2 I .
\end{equation}
Promoting \(q\) to the coordinate-register operator \(\hat q\) gives
\begin{equation}
\mathcal H_{\rm SM}(\hat q)
=
I\otimes H_{\rm SM}
-\kappa \hat q\otimes Q_{20}
+\frac{C}{2}\hat q^2\otimes I .
\end{equation}
Realizing the below-incumbent oracle for this family requires sufficiently
accurate coherent ground-state access and spectral comparison.

This architectural example introduces \(q\) through an external quadrupole
field; its numerical realization would require the ground-state access and
spectral comparison described above. The same construction
extends to multidimensional families, such as quadrupole--octupole fields
\((q_2,q_3)\) or Cartesian quadrupole components
\((q_{20},q_{22})\), provided their parameter dependence and the required
oracle ingredients remain implementable.

\section{QDMin minimum-search layer}

Grover diffusion uses the threshold phase in
Eq.~\eqref{eq:general-threshold-oracle} to increase the probability of
measuring a coordinate with energy below the incumbent \(E_t\). Repeating
this search with successively lower thresholds implements minimum finding.

For a known number of marked candidates, repeated Grover iterations amplify
the marked subspace~\cite{grover1996search}; amplitude amplification gives the
corresponding general framework~\cite{brassard2002amplitude}. In QDMin, however,
the number of coordinates below the incumbent is not known in advance and
changes whenever the incumbent is updated. We therefore use the
Boyer--Brassard--H{\o}yer--Tapp (BBHT) strategy, which draws the number of
Grover iterations randomly from a range that grows geometrically
~\cite{boyer1998tight}. D\"urr--H{\o}yer minimum finding embeds this search in
an incumbent-update loop: when a lower-energy coordinate is measured, it
becomes the new incumbent and defines a stricter threshold for the next
round~\cite{durr1996minimum}.

At the abstract query level, this procedure has expected
\(O(\sqrt{N_q})\) candidate-query complexity. This square-root behavior is the
standard D\"urr--H{\o}yer result inherited by QDMin. The bound counts calls
to the many-body threshold oracle; its gate-level implementation cost is
discussed separately below.

In the structured implementation, the search begins from a seeded random
candidate, which is taken as the initial incumbent. Its energy is estimated
using 63 shots of the same structured-QPE access circuit used throughout the
algorithm, and the modal QPE bin defines the threshold \(E_t\). Each BBHT
round then proceeds in three steps:

\begin{enumerate}
\item construct the coherent threshold oracle for the current incumbent;
\item apply the randomized number of Grover iterations and measure a candidate
      coordinate;
\item independently estimate the measured candidate energy and replace the
      incumbent only if the measured QPE bin is lower.
\end{enumerate}

The diffusion operator is implemented directly on the coordinate register as
an \(H\)-\(X\)-multi-controlled-\(Z\)-\(X\)-\(H\) reflection.
The structured coherent-access and QPE circuits supply all candidate-energy
information used during the search.

Because both state preparation and energy estimation are approximate, the
threshold predicate is also imperfect. Bounded-error quantum search requires
additional analysis in general~\cite{hoyer2003boundederror}. We report the
realized finite-shot search trajectory and validate it after the returned
coordinate has been frozen. This tests the implemented
oracle-and-amplification mechanism; a fault-tolerant success guarantee
requires a quantitative bound on the oracle errors.

The overlap-robust test of Sec.~\ref{sec:robust-predicate} supplies such a
bound. On every branch whose ground energy is at least \(\varepsilon\) away
from the threshold, each call gives the correct sign up to an amplitude error
of \(2\delta^2\), or \(2P_L(\ell)\) on the unmarked side, at the cost of
\(2L\) preparations instead of two. Branches closer than \(\varepsilon\) to
the incumbent cannot be decided, so the search can only promise a coordinate
whose energy lies within about \(\varepsilon\) of the minimum. This floor is
set by the energy precision and applies to any minimum search that works by
comparing energies.

We count candidate-energy measurements, coherent threshold calls, Grover
iterations, and circuit shots separately. Reference-oracle calculations use
precomputed exact energy ranks to test the search schedule and its scaling;
the structured execution obtains its energies from circuits. Round and query
limits bound the execution cost. Global optimality is established by exact
validation after the returned coordinate has been fixed.

\section{Circuit-level validation on a schematic family}
\label{sec:validation}

This section executes the complete pipeline---coherent encoding, ground-state
preparation, structured QPE, reversible comparison, uncomputation, and
search---at the circuit level on a four-orbital pairing-plus-field family
small enough for exact validation. Realistic landscapes follow in
Secs.~\ref{sec:sd-shell} and~\ref{sec:hubbard}.

\subsection{Model, encoding, and physical benchmark}

Pairing and collective polarization compete in the four-orbital benchmark.
The field favors occupation of one Kramers pair because the two pairs have
opposite quadrupole-like moments, whereas pairing mixes their configurations.
The model uses single-particle energies \((-0.6,-0.6,0.6,0.6)\), moments
\((-1,-1,1,1)\), pairing strength \(G=0.7\), collective-field coupling
\(\kappa=0.8\), and restoring strength \(C=2.1\).

For the 16-point grid
\[
q_k=-1.2+0.16k,\qquad k=0,\ldots,15,
\]
used to validate the coherent Hamiltonian construction, every direct
many-body Hamiltonian \(H(q_k)\) agrees with the corresponding coordinate
block of \(\mathcal H(\hat q)\) to double-precision accuracy. The exact
fixed-particle-number spectrum has its lowest grid energy at index 3,
corresponding to
\[
q=-0.72,\qquad E_0=-2.6096368,
\]
with both ends of the interval lying at higher energy. These exact
diagonalizations provide the validation benchmark for the circuit-generated
search result.

The continuous minimum lies near \(q=-0.731\), where the induced polarization
satisfies the stationary field--response balance in
Eq.~\eqref{eq:stationary-field-balance}. The energy minimum therefore reflects
the balance between the restoring field and the interacting many-body
response.

The primary dynamic search uses the four-point subset
\[
(-0.88,-0.72,-0.56,-0.40).
\]
Its compact coherent Hamiltonian contains 26 Pauli strings, increasing to 33
for the eight-point extension. Section III derives these counts without
enumerating coordinate projectors.

\subsection{Ground-state access and spectral accuracy}

The same adiabatic schedule prepares every branch with fidelity above
\(0.996\) on both the four- and eight-point one-dimensional grids
(Table~\ref{tab:prep-extension}). The coordinate labels are preserved to
numerical accuracy, and preparation followed by its implemented inverse
returns the input state to statevector precision. The accessible eigenphases
remain within the certified interval, so spectral wrapping does not affect
the energy comparison. Exact eigenstates and spectra enter these checks only
after the search result has been fixed.

\begin{table*}[tbp]
  \caption{Ground-state preparation and spectral validation. Qubit counts are
  given as coordinate, physical, and phase/work registers. \(F_{\min}\) and
  \(F_{\rm mean}\) denote the minimum and mean branch ground-state fidelities,
  respectively. CX and depth refer to one ASP circuit transpiled to
  \(\{R_z,\sqrt X,X,{\rm CX}\}\). The eight-point row is a one-dimensional
  extension test, while the 16-candidate row corresponds to the structured
  \(4\times4\) two-coordinate execution discussed below.}
  \label{tab:prep-extension}
  \begin{ruledtabular}
  \begin{tabular}{cccccccc}
  \(N_q\) & \((n_q,n_{\rm p},m)\) & Pauli strings &
  \(F_{\min}\) & \(F_{\rm mean}\) & min. gap & ASP CX/depth &
  certified \([E_L,E_U]\) \\
  \hline
  4 & (2,4,9) & 26 & 0.997870 & 0.998922 & 1.2 & 944/1362 &
  [-3.20952,3.42112] \\
  8 & (3,4,9) & 33 & 0.996836 & 0.998017 & 1.2 & 1072/1410 &
  [-3.20952,4.63200] \\
  16 & (4,4,10) & 41 & 0.744393 & 0.921122 & 1.2 & 1936/2938 &
  [-7.30950,8.37850] \\
  \end{tabular}
  \end{ruledtabular}
\end{table*}

The nine-qubit phase register resolves energy bins of about \(0.015\) in
model units. In the eight-point extension, the modal estimate at \(q=-0.72\)
lies within one bin of the postvalidated ground energy.

\subsection{Dynamic QDMin with circuit-generated energies}

\begin{figure*}[tbp]
  \centering
  \includegraphics[width=\textwidth]{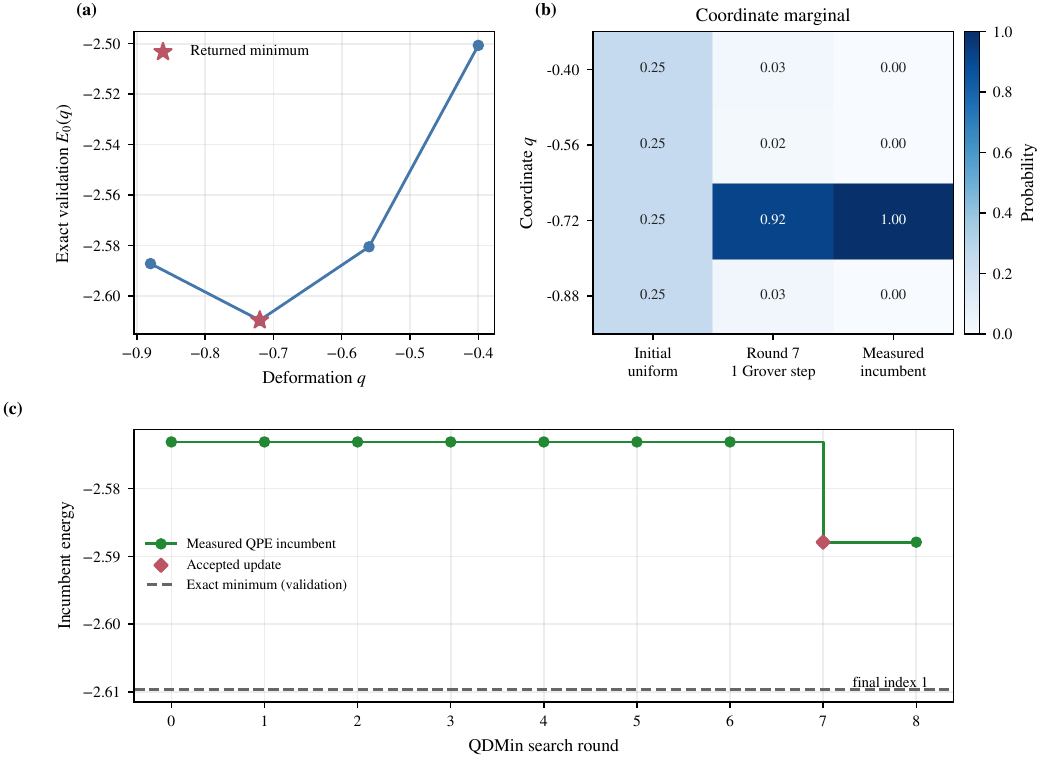}
  \caption{Dynamic four-point QDMin with circuit-generated energies.
  (a) Exact candidate energies are added only after the returned result has
  been frozen. (b) The coordinate distribution begins uniformly at
  probability \(0.25\) per candidate and is concentrated on index 1 by the
  single nonzero-Grover round. (c) The incumbent is updated only when an
  independently measured QPE bin is lower. Search rounds count algorithmic
  iterations.}
  \label{fig:dynamic-trajectory}
\end{figure*}

One Grover iteration raises the probability of the true minimum from
\(0.25\) to \(0.921\) in the four-point calculation
(Fig.~\ref{fig:dynamic-trajectory}). The amplified candidate, index 1, is then
measured and accepted through an independent QPE energy estimate. The
many-body threshold oracle therefore amplifies the minimum using energies
generated by the circuit.

The search receives the analytic Hamiltonian, grid, spectral certificate, and
known initial determinant. Circuit execution supplies the energy estimates;
its return is recorded before exact diagonalization, which subsequently
confirms \(q^\star=-0.72\) and \(E_0=-2.6096368\).
Integration tests verify this separation with repository eigensolvers
disabled.

With seed 23, the search begins at index 0. The first six BBHT rounds draw
zero Grover iterations; round 7 draws the single amplifying iteration. The
candidate at index 1 has a lower modal QPE energy than the incumbent and
remains the incumbent through the eight-round limit. The work registers return
cleanly with probability \(0.903\). This single seeded trajectory demonstrates
the oracle-and-amplification mechanism; its stopping limit supplies no
independent certificate of global optimality.

\subsection{Accuracy of structured energy access}

\begin{figure*}[tbp]
  \centering
  \includegraphics[width=\textwidth]{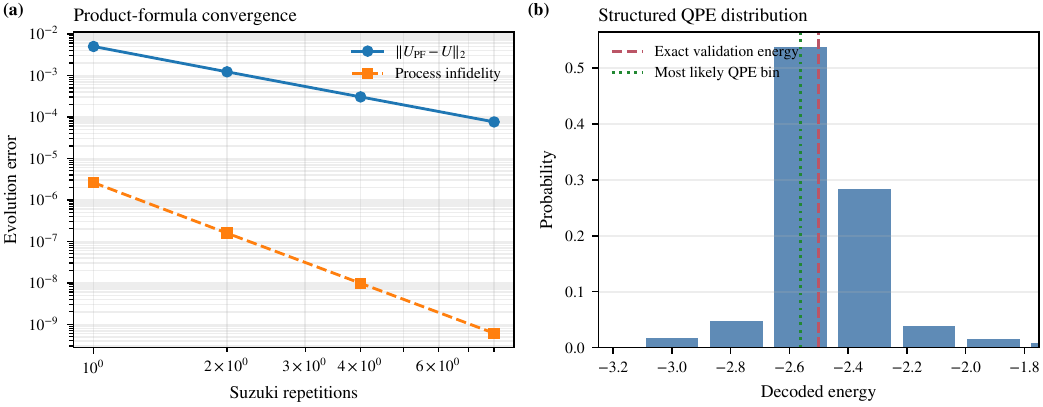}
  \caption{Validation of structured many-body energy access.
  (a) The second-order Suzuki error decreases systematically with repetition
  count. (b) A separate five-bit structured-QPE test resolves the validation
  energy within the expected phase-bin precision. The vertical lines mark the
  exact energy and the most-probable decoded QPE bin.}
  \label{fig:structured-validation}
\end{figure*}

The structured evolution preserves the coordinate blocks while its
spectral-norm error decreases with increasing Suzuki repetition count
(Fig.~\ref{fig:structured-validation}). A separate five-bit calculation
resolves a validation energy within half a phase bin and permits inverse-QPE
uncomputation to numerical accuracy. These checks characterize the
energy-access circuit independently of the minimum-search outcome, using
exact-state and dense-evolution references for validation.
Appendix~\ref{app:checks} gives the detailed residuals, spectral checks, and
execution records.

\subsection{Two-coordinate execution and preparation limits}

The complete circuit also operates on two coherent collective coordinates.
The second field couples through the number-conserving transition operator
\begin{equation}
Q_3
=
\sum_{s=0}^{1}
\left(
c_s^\dagger c_{s+2}
+\mathrm{H.c.}
\right),
\end{equation}
with coupling \(\kappa_3=1.0\) and restoring strength \(C_3=1.2\). Within this
schematic four-orbital model, \(Q_3\) defines an octupole-like transition
field. A calibrated nuclear octupole operator would require specified orbital
parities, angular momenta, and physical multipole matrix elements.
Two coordinate qubits encode four quadrupole-field values \(q_2\), and two
more encode four octupole-field values \(q_3\), producing a
\(4\times4\) grid of 16 candidate Hamiltonians.

The 41-string coherent Hamiltonian reproduces all 16 coordinate blocks to
numerical accuracy. Its spectral certificate covers the continuous coordinate
rectangle (Table~\ref{tab:prep-extension}), and the ten-qubit phase register
resolves energy bins of about \(0.018\). Post-search spectra confirm that no
accessible eigenphase wraps across the phase boundary.

With seed 223607, the search returns index 9 after 16 rounds and 23 combined
energy and threshold queries. Exact diagonalization, performed after the
return is fixed, identifies the same unique minimum,
\[
(q_2,q_3)=(-0.5,1.2),
\qquad
E_0=-2.7750621,
\]
with a structured energy estimate about \(0.018\) above the exact value.

\begin{figure*}[tbp]
  \centering
  \includegraphics[width=\textwidth]{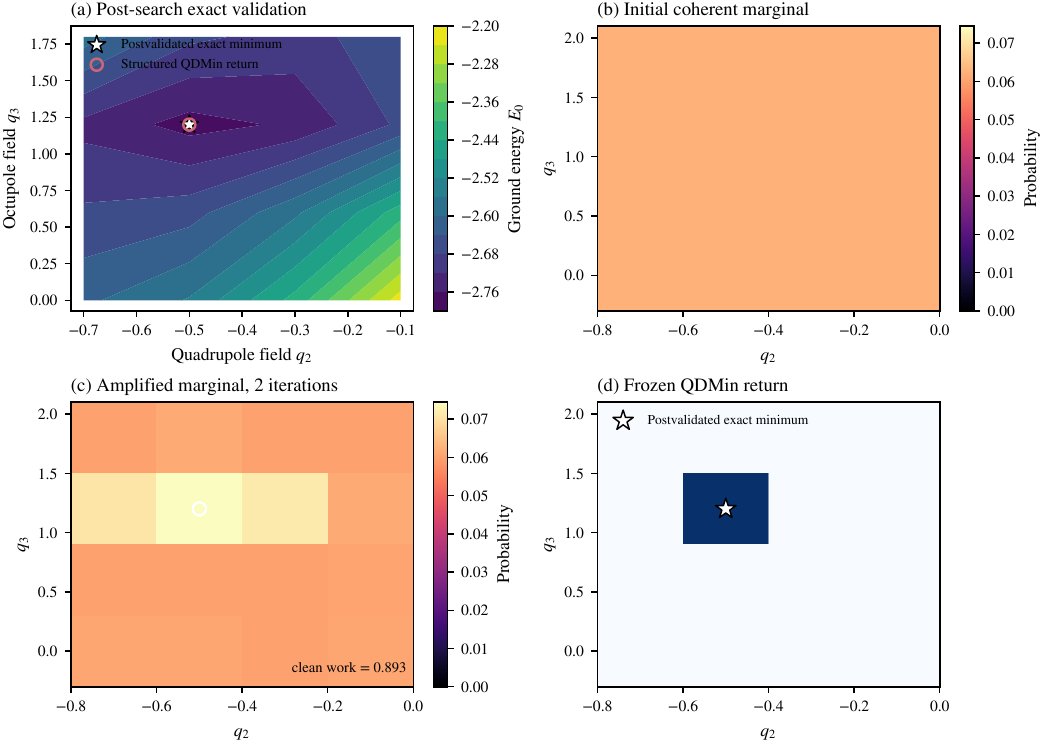}
  \caption{Two-coordinate execution on a \(4\times4\) grid with
  circuit-generated energy estimates.
  (a) Exact energies are introduced only after the structured return has been
  fixed. (b) Before amplification, each of the 16 coordinate candidates has
  uniform probability \(1/16=0.0625\). (c) A representative two-iteration
  Grover round executes the complete structured threshold, uncomputation, and
  diffusion sequence after the final incumbent has already been sampled.
  The work registers return cleanly with probability \(0.893\). (d) Post-search
  exact diagonalization confirms the unique grid minimum. The coordinate axes
  represent the schematic quadrupole- and octupole-field coordinates.}
  \label{fig:two-dimensional}
\end{figure*}

Starting from index 14, the search first encounters index 9 in
round 5 with zero Grover iterations, when each candidate has uniform
probability \(1/16\). Later rounds execute the full multidimensional
threshold and diffusion sequence. A representative two-iteration round raises
the probability of index 9 from \(0.0625\) to \(0.07445\), with clean-work
probability \(0.893\) (Fig.~\ref{fig:two-dimensional}). The 2D calculation
thus demonstrates circuit integration; the 1D calculation demonstrates
amplitude-assisted minimum selection.

This run reached the right answer but missed its own preparation target:
the worst branch fidelity was \(0.744\), against a declared \(0.98\)
(Table~\ref{tab:prep-extension}). The rest of this section explains why,
repairs it, and replaces the single trajectory by statistics over many
seeded runs.

\subsection{Repairing preparation and precision}
\label{sec:validation-repair}

The fidelity loss is not caused by a small gap: along every branch's
preparation path the gap stays at or above \(1.2\). The cause is the linear
ramp \(s(u)=u\), which switches the interaction on abruptly. The error it
leaves behind falls only as \(1/T\), and the product-formula error of the
finite time steps adds to it. A ramp that starts and ends with zero slope,
\(s(u)=3u^2-2u^3\), removes the leading error without any knowledge of the
gap. With total time \(T=12\) and 48 time steps, every branch reaches
fidelity \(0.988\) or better.

Energy precision was a second limit, hidden behind the first. Each phase bit
of the original QPE used two product-formula repetitions per unit of
evolution. That shifts the measured energies by up to \(0.036\), two phase
bins, while the two lowest grid energies differ by only \(0.021\). Eight
repetitions reduce the shift to \(0.005\) with eleven phase qubits and
\(0.004\) with twelve. The uniform phase register still leaks about
\(5\%\) of a component's weight across a threshold two bins away; a Kaiser
window reduces this to about \(3\times10^{-5}\).

Table~\ref{tab:robust-search} summarizes 2000 seeded searches for each
setting, all with energies generated by the circuits. They were simulated by
applying the exact operators of the same circuits, which reproduces the
gate-level simulation to \(10^{-8}\) at a small fraction of its cost.
Three observations stand out.

First, the robust test does what it is designed to do. With the old
preparation, the plain oracle flips some branches only halfway, an error of
\(0.51\) in amplitude; the fixed-point version reduces this to \(0.020\).

Second, on 16 points this hardly matters for the outcome. With fine enough
phase resolution, both oracles find the minimum in \(99.4\%\) of runs, even
with the old \(0.744\) preparation. A search on 16 points makes only about
eight coherent calls, and every candidate it proposes is checked by a
separate 63-shot energy measurement, so occasional oracle errors are caught.

Third, the threshold has to sit clear of the energies being compared. The
coherent threshold is placed two bins below the incumbent's measured energy,
so that the incumbent itself is not half-marked. With eleven phase qubits the
true minimum then lies just below the edge of the unresolved window: the
plain oracle half-marks it and succeeds in \(95\%\) of runs, whereas the
fixed-point oracle completes the mark and succeeds in \(99\%\). With twelve
phase qubits the window is narrow enough for both.

\begin{table*}[tbp]
\caption{Dynamic QDMin on the two-coordinate model with circuit-generated
energies, 2000 seeded runs per entry (standard errors \(0.002\)--\(0.005\)).
QPE settings: phase qubits \(m\), product-formula repetitions per base
power, and phase window. ``Margin'' is the distance, in bins, of the coherent
threshold below the incumbent's measured energy. ``Ramp'' is the
preparation schedule, linear (\(T=4\), 8 steps) or smoothstep (\(T=12\), 48
steps), and \(F_{\min}\) its worst branch fidelity. ``Oracle error'' is the largest deviation of a
branch amplitude from \(\pm1\) over thresholds at least two bins from every
grid energy. ``Success'' is the probability of returning the global minimum;
for the \(8\times8\) grid, whose two lowest energies differ by \(0.0006\),
far below a bin, it is the probability of returning a point within two bins
(\(0.017\)) of it.}
\label{tab:robust-search}
\begin{ruledtabular}
\begin{tabular}{llcccc}
Grid & \(m\), reps, window & Margin & Ramp (\(F_{\min}\)) &
Oracle error & Success \\
 & & & & plain / robust & plain / robust \\
\hline
\(4\times4\) & 10, 2, uniform & 0 & linear (0.744) & 0.50 / 0.10 & 0.991 / 0.974 \\
\(4\times4\) & 11, 8, Kaiser & 0 & linear (0.744) & 0.51 / 0.020 & 0.989 / 0.971 \\
\(4\times4\) & 11, 8, Kaiser & 0 & smooth (0.988) & 0.024 / 0.007 & 0.990 / 0.970 \\
\(4\times4\) & 11, 8, Kaiser & 2 & linear (0.744) & 0.51 / 0.020 & 0.948 / 0.988 \\
\(4\times4\) & 11, 8, Kaiser & 2 & smooth (0.988) & 0.024 / 0.007 & 0.958 / 0.990 \\
\(4\times4\) & 12, 8, Kaiser & 2 & linear (0.744) & 0.51 / 0.020 & 0.994 / 0.995 \\
\(4\times4\) & 12, 8, Kaiser & 2 & smooth (0.988) & 0.024 / 0.007 & 0.994 / 0.994 \\
\(8\times8\) & 11, 8, Kaiser & 2 & smooth (0.984) & 0.031 / 0.009 & 1.000 / 1.000 \\
\end{tabular}
\end{ruledtabular}
\end{table*}

The \(8\times8\) grid shows where the robust test starts to pay. Its two
lowest energies differ by \(0.0006\), far below the bin width of
\(0.0086\), so no comparison at this precision can separate them, and the
search can only promise a point close to the minimum. Both oracles return a
point within two bins of the minimum in every one of the 2000 runs, using
about 31 coherent calls, or \(3.9\sqrt{64}\). The robust oracle returns
better points on average: the mean energy error of the returned point is
\(0.0017\), half the plain oracle's \(0.0034\). It also leaves the work
registers much cleaner. In the worst search round simulated, the plain
oracle returns them clean with probability only \(0.10\), the robust oracle
with \(0.39\). Errors per call accumulate over the number of calls, so this
difference is expected to grow with the grid.

The same field--response interpretation used in one dimension extends to the
two-coordinate family. At an interior stationary point,
\begin{equation}
C_2q_2=\kappa_2\langle Q_2\rangle,
\qquad
C_3q_3=\kappa_3\langle Q_3\rangle .
\end{equation}
The chosen nonnegative \(q_3\) values define the controlled two-coordinate
benchmark.

\subsection{Classical optimization on the schematic landscape}

Exhaustive evaluation is the cheapest tested classical procedure on this
small grid. It uses 16 energy evaluations, compared with 65 for single-start
Nelder--Mead and 142 for three-start Nelder--Mead
~\cite{neldermead1965simplex}, and 186 for seeded differential
evolution~\cite{stornprice1997de}. All continuous
optimizers converge near \(q=-0.73068\), consistent with the continuous
minimum above.

The agreement shows that this landscape contains smoothness that classical
methods can exploit. The evaluation counts characterize the small benchmark.
A total-cost advantage for expensive microscopic Hamiltonians remains to be
tested against methods that exploit this classical structure.

\section{Realistic landscapes: static quadrupole fields in the \textit{sd} shell}
\label{sec:sd-shell}

The schematic model of Sec.~\ref{sec:validation} tests the circuits; it
cannot tell us whether the search helps on the landscapes that a realistic
interaction produces. For that we turn to the \textit{sd} shell, where a
well-tested effective interaction is available and the nuclei near
\(^{24}\)Mg are known to be deformed. Here we study the search layer only: the
energies come from exact diagonalization and stand in for the oracle, so the
question is how many energy queries the search needs, not how they are
produced.

The coherent oracle cannot be simulated at this size. The \textit{sd} shell
has 24 single-particle states, so the nucleus alone needs 24 qubits; a
five-channel field with 16 values per channel adds 20 coordinate qubits, and
resolving the energy adds about a dozen phase qubits. A circuit of some 56
qubits is far beyond classical simulation, which is precisely the regime in
which a quantum oracle would be useful. We therefore separate the two
questions. Section~\ref{sec:validation} shows, on a model small enough to
simulate, that the oracle works and what it costs; this section uses the
oracle's ideal output, the exact ground energies, to measure how many queries
the search needs on realistic landscapes. The query counts are the factor
that multiplies the cost of each oracle call; that cost itself is not part
of this comparison.

\subsection{The static quadrupole family}

We use the USDB interaction~\cite{brown2006usd} for nucleons in the
\(0d_{5/2}\), \(1s_{1/2}\), and \(0d_{3/2}\) orbits above an \(^{16}\)O core.
The parametrized family is the static-field decoupling of
Sec.~\ref{sec:problem} applied to the quadrupole operator: a static field
\(\bm q\) with five components, one for each Cartesian component \(A_a\) of
the mass quadrupole operator \(Q_{2\mu}=\sum_ir_i^2Y_{2\mu}(\hat r_i)\),
\begin{equation}
H(\bm q)=H_{\rm USDB}+\frac{\chi}{2}\sum_{a=1}^{5}\left(A_a-q_a\right)^2 .
\label{eq:sd-family}
\end{equation}
The penalty term pulls the quadrupole moment of the nucleus towards the
field \(\bm q\), with stiffness \(\chi\). Every member lies above the USDB
ground energy, and at the minimum the field equals the nucleus's own intrinsic
quadrupole moment (Sec.~\ref{sec:problem}). Because the interaction is
rotationally invariant, the energy depends only on the shape of the field,
not on its orientation. The five components therefore reduce to the familiar
deformation \(\beta\) and triaxiality \(\gamma\), shown as the point
\((q_{20},q_{22c})=(\beta\cos\gamma,\beta\sin\gamma)\) in the intrinsic plane.
Fields are measured in units of \(b^2\), with \(b\) the oscillator length.

Each energy is obtained by Lanczos diagonalization in the full shell-model
space, up to about \(2.5\times10^5\) states for \(^{24}\)Mg. As a check, the
implementation reproduces published USDB results for \(^{24}\)Mg: the first
\(2^+\) state at \(1.502\) MeV, \(B(E2)=95\) e\(^2\)fm\(^4\), and
\(Q(2^+_1)=-19.35\) e\,fm\(^2\)~\cite{jokiniemi2023muoncapture}. The signs of
the spectroscopic quadrupole moments match the expected shapes: prolate for
\(^{20}\)Ne and \(^{24}\)Mg, oblate for \(^{28}\)Si.

\subsection{Shapes selected by the static field}

\begin{figure*}[tbp]
  \centering
  \includegraphics[width=0.96\textwidth]{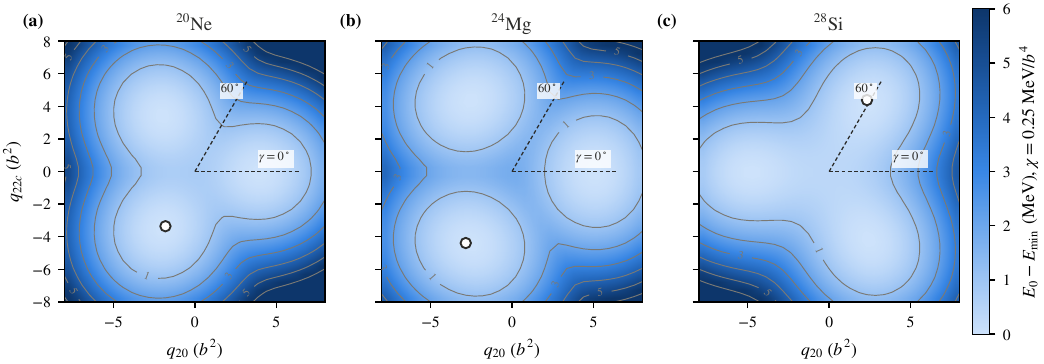}
  \caption{Static quadrupole-field landscapes \(E_0(q_{20},q_{22c})-E_{\min}\)
  for USDB at \(\chi=0.25\) MeV\(/b^4\), from exact Lanczos diagonalization
  on \(32\times32\) grids. Dashed lines mark \(\gamma=0^\circ\) (prolate) and
  \(\gamma=60^\circ\) (oblate); the plane is symmetric under
  \(\gamma\to\gamma\pm120^\circ\), so each minimum appears three times. Open
  circles mark the lowest grid point: a symmetry image of the prolate
  minimum for \(^{20}\)Ne and \(^{24}\)Mg, and the oblate minimum for
  \(^{28}\)Si. Contours every 1 MeV.}
  \label{fig:sd-landscapes}
\end{figure*}

The stiffness \(\chi\) decides whether a deformed minimum appears at all. A
weak field cannot pay for the correlation energy that deformation costs, and
the minimum stays at \(\bm q=0\). Above a threshold that depends on the
nucleus, the minimum moves to a deformed, self-consistent field. At
\(\chi=0.25\) MeV\(/b^4\) all three nuclei are deformed with their observed
shapes: \(^{20}\)Ne and \(^{24}\)Mg are prolate, with oblate local minima,
and \(^{28}\)Si is oblate (Fig.~\ref{fig:sd-landscapes}). Each minimum
appears three times in the plane because rotating the shape by \(120^\circ\)
in \(\gamma\) describes the same nucleus.

\subsection{Search on realistic landscapes}

We compare three ways of finding the lowest grid point, counting energy
queries until a point within 1 keV of the minimum is first found: QDMin
with an ideal oracle, multistart steepest descent on the grid, which
exploits smoothness, and a random-order scan. A QDMin query costs about
\(2\xi\) pointwise solves [Eq.~\eqref{eq:pointwise-crossover}], whereas the
classical counts are pointwise solves already. Table~\ref{tab:sd-search}
lists the results.

The outcome depends on the landscape, not on its size. On the
\(32\times32\) intrinsic planes the landscapes are smooth, and descent needs
about half as many queries as QDMin. On five-component grids with 16 values
per channel, about \(10^6\) points, both need a few hundred to a thousand
queries, one to two orders of magnitude fewer than a random scan, and on the
single-well landscapes they are comparable. The difference appears when the
physics provides competing minima. For \(^{28}\)Si at a larger stiffness,
\(\chi=2\), an oblate global minimum coexists with a prolate local minimum
6 MeV higher. Descent is repeatedly trapped in the prolate well and needs
about four times as many queries as QDMin on average, and eight times as
many in the worst 1\% of runs. Counted in pointwise solves, QDMin is cheaper
on this landscape if \(\xi\lesssim2\).

\begin{table*}[tbp]
\caption{Energy queries until a grid point within 1 keV of the global
minimum is first obtained: mean (99th percentile) over 2000--4000 seeded
runs. QDMin counts ideal threshold calls plus energy reads; descent and
random scan count pointwise solves. Five-component grids have
\(16^5\approx10^6\) points with fields in \([-4.4,4.4]\,b^2\), built by
bicubic interpolation of exact \((\beta,\gamma)\) landscapes; checked
against exact values at random grid points, the interpolation error is at
most 0.7 keV for \(^{20}\)Ne and \(^{24}\)Mg, 0.03 keV for \(^{28}\)Si at
\(\chi=0.25\), and 55 keV (0.1 keV at the minimum) for the steeper \(^{28}\)Si
landscape at \(\chi=2\), whose statistics therefore refer to the interpolated
landscape. \(\chi\) in MeV\(/b^4\).}
\label{tab:sd-search}
\begin{ruledtabular}
\begin{tabular}{lcrrrr}
Nucleus & \(\chi\) & \(N_q\) & QDMin & Descent & Random \\
\hline
\(^{20}\)Ne & 0.25 & 1024 & 75 (167) & 41 (132) & 341 (913) \\
\(^{24}\)Mg & 0.25 & 1024 & 75 (167) & 40 (126) & 347 (925) \\
\(^{28}\)Si & 0.25 & 1024 & 79 (190) & 44 (140) & 341 (917) \\
\(^{20}\)Ne & 0.25 & \(16^5\) & 560 (1381) & 476 (2094) & 31\,500 \\
\(^{24}\)Mg & 0.25 & \(16^5\) & 373 (922) & 405 (1865) & 12\,800 \\
\(^{28}\)Si & 0.25 & \(16^5\) & 419 (1040) & 326 (1350) & 14\,600 \\
\(^{28}\)Si & 2 & \(16^5\) & 1057 (2281) & 4146 (17\,571) & 118\,000 \\
\end{tabular}
\end{ruledtabular}
\end{table*}

Classical knowledge can also be combined with the quantum search. Instead
of starting from a uniform superposition, the search can start from a state
weighted towards points that a cheap classical estimate considers
promising. Amplitude amplification keeps these relative weights, so the
search still finds the true minimum, only faster when the estimate is good.
For \(^{28}\)Si at \(\chi=2\), an estimate built from 1024 exact energies on a
coarse sub-grid cuts the search to about 300 queries on average.

One caveat applies to all five-component results. Rotational invariance is
itself classical structure: it reduces the five-dimensional search to the
two-dimensional plane, where classical exploration is cheap. These grids are
therefore realistic-size tests of the search, not a case where QDMin beats a
classical calculation that uses the symmetry. A genuine test needs a family
whose dimension stays large after the symmetry is removed. Combined
quadrupole and octupole fields are an example (twelve channels, nine after
removing the orientation), but they require a model space that spans two
major shells, which we leave to future work.

\section{Second example: competing orders in a Hubbard ring}
\label{sec:hubbard}

Nothing in the construction is specific to nuclei. Any interaction that can
be written as a sum of squares of order parameters gives a static-field
family, and competing orders give competing minima. A small Hubbard ring
shows this.

We take a half-filled ring of \(L\) sites with hopping \(t\) and on-site
repulsion \(U\), and add infinite-range couplings in two competing channels:
the staggered magnetization \(A_s\), which favours a spin-density wave
(SDW), and the staggered charge \(A_c\), which favours a charge-density wave
(CDW),
\begin{align}
H&=H_{\rm Hub}-\frac{g_s}{2}A_s^2-\frac{g_c}{2}A_c^2,
\label{eq:hubbard-parent}\\
A_{s,c}&=\frac{1}{\sqrt L}\sum_i(-1)^i\left(n_{i\uparrow}\mp n_{i\downarrow}\right).
\end{align}
The infinite-range terms play the role of the couplings between chains in
quasi-one-dimensional materials, which are usually treated at mean-field
level, while the on-site repulsion \(U\) is kept exactly. Decoupling the two
squares gives the two-field family
\begin{equation}
H(q_s,q_c)=H_{\rm Hub}-g_sq_sA_s-g_cq_cA_c+\frac{g_s}{2}q_s^2+\frac{g_c}{2}q_c^2 ,
\label{eq:hubbard-family}
\end{equation}
an instance of Eq.~\eqref{eq:static-field-family}. As for a deformed
nucleus, the ground state of the finite ring is symmetric and has zero
order parameter. The order it would develop is visible only through the
family.

For a ten-site ring with \(U=4t\), the two orders behave differently. A
spin-ordered minimum appears already for weak coupling,
\(g_s\gtrsim0.3t\). A charge-ordered minimum appears only near
\(g_c\approx4.2t\), and then abruptly, because putting two electrons on one
site costs \(U\). With \(g_s=0.5t\) and \(g_c=4.4t\) the landscape has both:
SDW global minima at \(q_s\approx\pm2.5\), where the field matches the ring's
own staggered magnetization, and CDW local minima \(0.53t\) higher.

On a \(32\times32\) grid, finding a global minimum within \(10^{-3}t\) takes
41 ideal QDMin queries on average (103 in the worst 1\% of runs), against 52
(183) for multistart descent and 113 (421) for a random scan. As in
\(^{28}\)Si, the competing minimum traps local descent, and the advantage
shows up mainly in the tail.

\section{Resource accounting}
\label{sec:resources}

The cost of coherent many-body energy access determines whether the
candidate-query reduction is useful. We account separately for the work within
each query and its repetition by minimum finding.

Let
\[
\Delta E=\frac{E_U-E_L}{1-\eta}
\]
denote the affine energy range mapped to the QPE phase interval, and let
\(\epsilon\) be the desired energy-bin width. Binary
QPE~\cite{cleve1998algorithms} then requires
\begin{equation}
m=
\left\lceil
\log_2\left(\frac{\Delta E}{\epsilon}\right)
\right\rceil
\end{equation}
phase qubits.

A clean threshold oracle contains both forward and inverse energy access.
If the number of product-formula repetitions scales with the binary QPE
power, one forward QPE contains
\begin{equation}
r_{\rm base}(2^m-1)
\end{equation}
product-formula repetitions. The corresponding threshold cost can be written
schematically as
\begin{equation}
C_{\rm th}
=
2C_{\rm prep}
+
2C_{\rm QPE}(m)
+
C_{\rm cmp}(m),
\end{equation}
where \(C_{\rm prep}\) is the ground-state-access cost,
\(C_{\rm QPE}(m)\) is the \(m\)-bit phase-estimation cost, and
\(C_{\rm cmp}(m)\) is the reversible threshold-comparison cost.
 \(C_{\rm prep}=C_{\rm ASP}\) for the present demonstrator. The prefix
comparator requires at most \(m\) phase-flip blocks and introduces no flag or
carry ancillas.

Minimum finding also requires an energy measurement when a sampled coordinate
is tested as a possible new incumbent. Denoting this candidate-energy access
by
\begin{equation}
C_E
=
C_{\rm prep}
+
C_{\rm QPE}
+
C_{\rm read},
\end{equation}
a useful cost associated with one logical search opportunity is
\begin{equation}
C_{\rm access}
=
C_{\rm th}
+
C_E,
\end{equation}
up to schedule-dependent constants. The abstract query model assigns unit
cost to a threshold-oracle invocation; an end-to-end estimate includes
\(C_{\rm access}\).

\subsection{Comparison with a pointwise quantum scan}
\label{sec:pointwise-scan}

When \(E_0(\bm q)\) requires a quantum solver, the direct alternative to
QDMin is a pointwise quantum scan: \(N_q\) separate ground-energy
estimations at fixed parameters, each of cost \(C_E\), followed by a
classical minimum. This comparison isolates what the coherent register buys,
because both sides use the same many-body solver.

One threshold call computes the energy coherently, compares, and
uncomputes. Writing \(C_{\rm th}\approx2\xi C_E\), the factor
\(\xi\geq1\) collects the overhead of controlling every evolution on the
parameter register and of coherent rather than measured ground-state access.
If minimum finding uses \(a_{\rm th}\sqrt{N_q}\) threshold calls and
\(a_E\sqrt{N_q}\) energy reads, then
\begin{equation}
\frac{C_{\rm QDMin}}{C_{\rm scan}}
\approx
\frac{2\xi a_{\rm th}+a_E}{\sqrt{N_q}},
\quad
N_q^{\times}\approx\left(2\xi a_{\rm th}+a_E\right)^2 .
\label{eq:pointwise-crossover}
\end{equation}

The constants \(a_{\rm th}\) and \(a_E\) follow from an exact simulation of
the ideal BBHT/D\"urr--H{\o}yer schedule. With an ideal phase oracle, the
search state remains uniform on the marked and on the unmarked candidates,
so each round can be sampled exactly in \(O(1)\) classical work; grids up to
\(N_q=2^{24}\) were simulated with 4000 seeded runs per size. The mean number
of candidate queries until the global minimum is first measured is
\(3.0\)--\(3.4\sqrt{N_q}\) for \(2^{12}\leq N_q\leq2^{24}\), and a fixed
budget of \(5.3\sqrt{N_q}\) or \(7.4\sqrt{N_q}\) succeeds with probability
\(0.9\) or \(0.99\), respectively. These constants lie well below the
analytic bound \(22.5\sqrt{N_q}\)~\cite{durr1996minimum}. At
\(N_q=2^{24}\), threshold calls account for \(2.94\sqrt{N_q}\) and energy
reads for only \(0.10\sqrt{N_q}\) queries.

Bounding all queries by the 99\% budget gives
\(N_q^{\times}\lesssim220\,\xi^2\). The query count alone falls below
\(N_q\) only for \(N_q\gtrsim55\), so one-dimensional landscapes with a few
dozen points gain nothing. On a five-channel quadrupole grid with 16 points per
channel, \(N_q\approx10^6\), the ratio in
Eq.~\eqref{eq:pointwise-crossover} is \(0.014\xi\). Both approaches share
logarithmic error-control factors, and classical structure that reduces
\(N_q\), such as symmetry or smoothness, reduces it for both.

The comparison isolates what the coherent register contributes. Any advantage
of a quantum many-body solver over classical methods for a single
\(E_0(\bm q)\) is a property of the solver, and the pointwise scan shares it.
The two effects multiply: the classical crossover of
Eq.~\eqref{eq:crossover-condition} below contains the solver ratio
\(C_{\rm Q}/C_{\rm cl}\) and the search factor \(\alpha/\sqrt{N_q}\)
separately, and only the second is attributed to QDMin. In the regime that
motivates quantum simulation, where \(E_0(\bm q)\) cannot be obtained
classically at the required size, classical global-search strategies remain
available in hybrid form, driven by pointwise quantum energy estimates. The
relevant comparison then counts energy solves, as in
Table~\ref{tab:sd-search}.

The overlap-robust predicate of Sec.~\ref{sec:robust-predicate} enters
\(\xi\) as a multiplicative factor \(L=O(\log(1/\delta)/\sqrt{F_{\min}})\):
it buys a per-call error bound with \(2L\) instead of two applications of
preparation and QPE. It is worthwhile when the number of coherent calls is
large enough that uncontrolled per-call errors would accumulate, which is the
large-\(N_q\) regime in which QDMin is intended to pay.

The cost of coherent energy access need not be uniform over the grid:
preparation time grows as the path gap closes, and some branches may need
finer phase resolution than others. If branch \(q\) requires time \(t_q\),
a uniform oracle pays \(\max_qt_q\) per call. Variable-time search reduces
the search cost to \(O(\sqrt{\sum_qt_q^2})\) instead of
\(O(\sqrt{N_q}\max_qt_q)\)~\cite{ambainis2010variabletime}, which matters
when a few branches, such as those near level crossings, dominate the
worst-case cost. We do not implement it here.

\subsection{Scaling and potential quantum advantage}

Let \(L\) characterize the size of the many-body problem and
\(C_{\rm Q}(L)\) denote the cost of usable coherent energy access, including
the operations counted in \(C_{\rm access}\) above. The QDMin cost is
schematically
\begin{equation}
C_{\rm QDMin}
\sim
\alpha\sqrt{N_q}\,C_{\rm Q}(L),
\label{eq:qdmin-total-cost}
\end{equation}
where \(\alpha\) collects the schedule-dependent constants and repetitions of
bounded-error minimum finding.

For a like-for-like classical calculation of the same microscopic Hamiltonian
family, let \(C_{\rm cl}(L)\) denote the cost of obtaining one candidate
ground-state energy. Explicit evaluation of all \(N_q\) candidates then has
cost
\begin{equation}
C_{\rm exhaustive}
\sim
N_q\,C_{\rm cl}(L).
\label{eq:classical-total-cost}
\end{equation}
Their ratio is therefore
\begin{equation}
\frac{C_{\rm QDMin}}{C_{\rm exhaustive}}
\sim
\frac{\alpha}{\sqrt{N_q}}
\frac{C_{\rm Q}(L)}{C_{\rm cl}(L)}.
\label{eq:crossover-condition}
\end{equation}

The ratio \(C_{\rm Q}(L)/C_{\rm cl}(L)\) compares the many-body solvers;
the factor \(1/\sqrt{N_q}\) describes the reduction in candidate access. A
future quantum many-body solver advantage could therefore combine with the
minimum-search reduction. Testing their combined total-cost benefit
requires larger-system resource studies.

For the microscopic nuclear regime envisioned here, the appropriate
like-for-like comparison is with many-body calculations of the same
Hamiltonian, such as large-scale configuration-interaction
~\cite{caurier2005shellmodel} or no-core shell-model
calculations~\cite{barrett2013nocore} using realistic two- and three-nucleon
interactions.
Mean-field and energy-density-functional methods can provide substantially
cheaper descriptions of collective landscapes, but they solve a different
effective many-body problem and therefore address a different computational
comparison.

Whether both factors in Eq.~\eqref{eq:crossover-condition} become favorable
simultaneously remains problem dependent. For the adiabatic realization of
ground-state access used here, the preparation cost can increase strongly when
the minimum spectral gap along the path becomes small; other coherent
ground-state-access methods may have different resource tradeoffs.
Hamiltonian simulation, QPE precision, and fault-tolerant error correction
also contribute to \(C_{\rm Q}(L)\), while classical microscopic solvers can
exploit symmetries, sparsity, basis truncation, importance selection, and
other physical structure. The present calculations establish the
architecture and the resource quantities needed to evaluate this prospective
crossover.

\subsection{Structured implementation resources}

Resolving energy dominates the circuit width: the four-point calculation
uses 15 qubits, of which nine belong to the phase/work register, and the
original two-coordinate calculation uses 18 qubits, of which ten are
phase/work qubits. The repaired two-coordinate runs of
Sec.~\ref{sec:validation-repair} need eleven or twelve phase qubits, 19 or 20
qubits in total, because the comparison must resolve energy differences of
two to three phase bins. The ASP, window, and prefix-comparison circuits
require no additional ancillas. Table~\ref{tab:prep-extension} gives the ASP gate counts and depths.

Repeated energy access dominates the execution cost. The four-point
trajectory uses ten combined candidate queries, but its shot-level
implementation requires more than half a million spectral product-formula
repetitions. The two-coordinate trajectory uses 23 combined queries and more
than two million such repetitions (Table~\ref{tab:search-resources}). A
candidate-query count therefore omits most of the circuit work needed to
obtain each usable energy estimate.

\begin{table*}[tbp]
  \caption{Resources for the reported structured search trajectories.
  An energy batch contains 63 shots. Combined candidate queries count energy
  batches plus coherent threshold calls. Shots are counted separately.
  Preparation calls include the inverse preparations within threshold oracles.
  Product-formula counts refer to the implemented Suzuki sequences.}
  \label{tab:search-resources}
  \begin{ruledtabular}
  \begin{tabular}{lrr}
  Quantity & Four-point 1D & \(4\times4\) two-coordinate \\
  \hline
  Search rounds & 8 & 16 \\
  Candidate-energy batches & 9 & 17 \\
  Shot-level forward QPE executions & 567 & 1071 \\
  Coherent threshold calls & 1 & 6 \\
  Grover iterations & 1 & 6 \\
  Combined candidate queries & 10 & 23 \\
  Controlled Hamiltonian evolutions & 5121 & 10830 \\
  Spectral product-formula repetitions & 581518 & 2215818 \\
  Preparation and inverse-preparation calls & 569 & 1083 \\
  \end{tabular}
  \end{ruledtabular}
\end{table*}

One forward QPE requires 1022 product-formula repetitions in 1D and 2046 in
2D. Decomposing the reusable ten-bit 2D QPE stage to the chosen basis gives
about three million instructions at depth \(2.55\times10^6\).
Appendix~\ref{app:checks} gives the exact decomposition counts. Full native
transpilation of the nine-bit threshold oracle was not attempted because of
its synthesis cost; the separate five-bit, 11-qubit validation circuit has
67740 CX gates at depth 125354. These are small-circuit transpilation
resources; fault-tolerant estimates would also require error-correction
overhead.

\begin{figure*}[tbp]
  \centering
  \includegraphics[width=\textwidth]{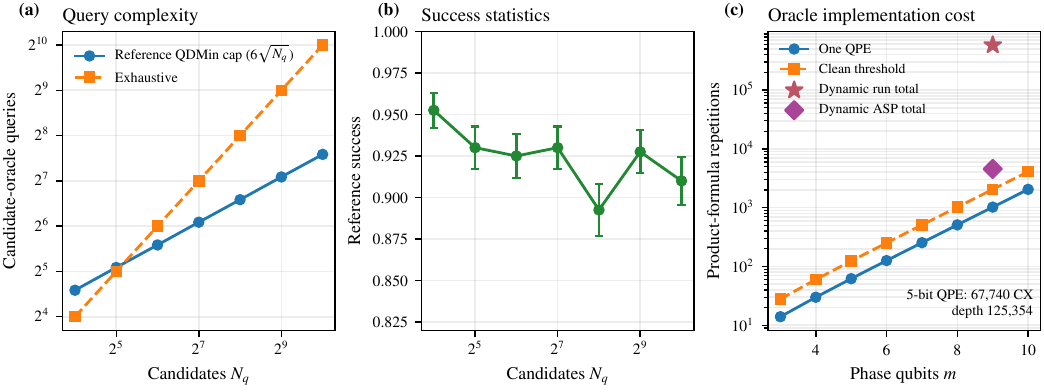}
  \caption{Candidate-query scaling and structured energy-access cost.
  (a) Candidate-query counts for the reference-oracle implementation under a
  configured \(6\sqrt{N_q}\) query cap, compared with exhaustive evaluation.
  (b) Success probabilities from 400 seeded reference-oracle trials at each
  candidate-space size. (c) Growth of the product-formula cost with QPE phase
  precision; the star marks the spectral contribution of the primary nine-bit
  dynamic calculation, the diamond marks its separately counted ASP
  contribution, and the annotation gives the transpiled resources of the
  standalone five-bit structured-QPE regression circuit. Panels (a) and (b)
  characterize the inherited minimum-search layer, whereas panel (c)
  characterizes the many-body energy-access layer.}
  \label{fig:scaling-resources}
\end{figure*}

Figure~\ref{fig:scaling-resources} separates the reference-oracle
candidate-query results in panels (a) and (b) from the structured
energy-access cost in panel (c). Increasing QPE precision raises the
product-formula cost even when the candidate-query count is unchanged.

For candidate spaces from \(N_q=16\) to \(1024\), the reference-oracle
calculations return the correct minimum with probabilities between 0.892 and
0.953 under the configured query cap. These trials verify that the implemented
randomized minimum-search schedule operates within the intended
square-root-scale candidate-query budget over the tested sizes. An
end-to-end speedup remains unestablished because these reference-oracle
trials isolate the search schedule from the many-body energy-access cost.

\section{Discussion}

QDMin connects two established ideas, coherent parameter registers and
quantum minimum finding, through an explicit many-body threshold oracle. The
circuit-level runs show that the connection works end to end: with energies
produced by the circuits themselves, the search finds the minimum of a
16-point two-coordinate landscape in \(99\%\) of runs, and a point within the
energy resolution of the minimum of a 64-point landscape in every run.

The runs also show which ingredients carry the weight. Ground-state access
looked like the bottleneck in our first two-coordinate calculation, but its
failure came from an abrupt ramp and coarse time steps, which a smooth,
finer ramp repairs without knowledge of the gap. Energy precision turned out to matter
as much: the comparison can only be as sharp as the phase estimate, and the
phase estimate needs both enough product-formula repetitions and a tapered
window. On small grids the search tolerates imperfect oracles, because it
makes few coherent calls and checks each candidate separately. The robust
test matters as the number of calls grows; on 64 points it already halves the
energy error of the returned point. For harder Hamiltonians, where preparation
fidelity is lower and grids are larger, the overlap-robust test is the
natural default, provided the phase estimate has suppressed tails.

The schematic model fixes the structure of the calculation, not its physical
content. The \textit{sd}-shell family of Sec.~\ref{sec:sd-shell} supplies
the physical content with a realistic interaction; running it at circuit
level would require preparation, simulation, and spectral bounds for that
model space, which we leave to future work.

QDMin is intended for families whose ground energies are themselves hard to
compute classically, and in that regime two potential gains combine. A
quantum solver may estimate a single \(E_0(\bm q)\) more cheaply than a
classical many-body method; QDMin then reduces the number of such estimates
from the \(N_q\) of a pointwise scan to \(O(\sqrt{N_q})\). The first gain is
not established in general. Its size depends on preparing states with
sufficient ground-state overlap, and a recent assessment finds no evidence
for a generic exponential quantum advantage in ground-state quantum chemistry,
while leaving room for polynomial speedups~\cite{lee2023evaluating}. QDMin
also asks more of ground-state access than a pointwise scan does: it requires
adequate preparation on every parameter branch simultaneously within one
coherent circuit, without branch-specific measurement or repetition. That
requirement is the factor \(\xi\) in Eq.~\eqref{eq:pointwise-crossover}. The
two-coordinate model shows it concretely: a ramp that sufficed for one
coordinate had to be made smoother and three times longer before every
branch was prepared well.
The \textit{sd}-shell landscapes of Sec.~\ref{sec:sd-shell} are classically
tractable; they test the search layer on realistic energy surfaces, while
the solver advantage becomes relevant only for model spaces well beyond
direct diagonalization. Within that regime, the comparison of
Table~\ref{tab:sd-search} indicates where the search factor matters: on
smooth single-well landscapes, a classical search strategy driven by
pointwise energies needs a comparable number of solves, whereas competing
minima, as in the oblate--prolate case of \(^{28}\)Si, favour the coherent
search.

A total-cost comparison must use the same physical Hamiltonian and comparable
energy accuracy. Configuration-interaction or no-core shell-model
calculations provide microscopic references, while mean-field and
energy-density-functional methods describe a different effective many-body
problem. Even for the same Hamiltonian, continuation, gradients, interpolation,
basin hopping, surrogates, and physics-informed initialization can reduce
classical work~\cite{walesdoye1997basinhopping,jones1998ego}. The smooth
one-dimensional benchmark illustrates the need for structure-aware
classical baselines alongside exhaustive enumeration. The crossover in
Eq.~\eqref{eq:crossover-condition} must be tested against the
strongest available classical treatment, including its use of physical
structure.

\section{Conclusion}

QDMin finds the lowest-energy member of a parametrized many-body family with
\(O(\sqrt{N_q})\) coherent oracle calls instead of the \(N_q\) solves of a
pointwise scan, by holding all candidates in a coherent register and
searching them with a ground-energy threshold oracle. Static-field families give the problem
a natural home: their minimum is a rigorous variational bound that the parent
ground state does not reveal. Because imperfect preparation can make the
threshold test miss a branch but never accept a wrong one, fixed-point
amplification makes the oracle robust at a cost logarithmic in the target
error and inverse-square-root in the preparation fidelity, provided the
phase estimate has suppressed tails.

The gain is real but conditional. It concerns the number of energy queries,
not the solver itself; it requires grids larger than about \(220\,\xi^2\) points; and
on realistic shell-model landscapes it appears where competing minima trap
local descent, not where the landscape is smooth. Circuit-level runs with
circuit-generated energies confirm that the complete pipeline finds the
minimum, and that preparation ramps and energy precision, rather than the
search itself, decide how well it does. The open question is the one that
decides any quantum many-body application: whether coherent ground-state
access for large, realistic Hamiltonians can be made cheap enough for the
search advantage to be collected.

\section*{Code and Data Availability}
Secondary numerical checks and detailed circuit counts are given in the
appendices. The source code, configurations, processed results, tests, and
figure scripts that reproduce every number and figure in this paper will be
released under the Apache-2.0 license, with an archived DOI, on acceptance.
During review they are available to the editors and referees, anonymously
and free of charge, on request. The USDB interaction is the publicly
available file in KSHELL format, fetched and checksum-verified by a script in
the repository; no experimental data were used.

\section*{Acknowledgments}
This work was supported by the Science and Engineering Research Board under Grant No. CRG/2022/009359.

\bibliographystyle{quantum}
\bibliography{references}

\begin{thebibliography}{10}

\bibitem{bender2003selfconsistent}
Michael Bender, Paul-Henri Heenen, and Paul-Gerhard Reinhard.
\newblock ``Self-consistent mean-field models for nuclear structure''.
\newblock \href{https://dx.doi.org/10.1103/RevModPhys.75.121}{Reviews of Modern
  Physics {\bf 75}, 121--180}~(2003).

\bibitem{heydewood2011coexistence}
Kris Heyde and John~L. Wood.
\newblock ``Shape coexistence in atomic nuclei''.
\newblock \href{https://dx.doi.org/10.1103/RevModPhys.83.1467}{Reviews of
  Modern Physics {\bf 83}, 1467--1521}~(2011).

\bibitem{butler2016octupole}
Peter~A. Butler.
\newblock ``Octupole collectivity in nuclei''.
\newblock \href{https://dx.doi.org/10.1088/0954-3899/43/7/073002}{Journal of
  Physics G: Nuclear and Particle Physics {\bf 43}, 073002}~(2016).

\bibitem{schunckrobledo2016fission}
Nicolas Schunck and Luis~M. Robledo.
\newblock ``Microscopic theory of nuclear fission: a review''.
\newblock \href{https://dx.doi.org/10.1088/0034-4885/79/11/116301}{Reports on
  Progress in Physics {\bf 79}, 116301}~(2016).

\bibitem{xiang2026triaxialquadrupoleoctupole}
J.~Xiang, J.~Zhao, Z.~P. Li, and D.~Vretenar.
\newblock ``Microscopic triaxial quadrupole-octupole collective {Hamiltonian}
  for low-energy nuclear excitations''.
\newblock \href{https://dx.doi.org/10.1103/xbqz-9tjj}{Physical Review C {\bf
  113}, 044328}~(2026).

\bibitem{dobaczewski2021hfodd}
J.~Dobaczewski, P.~B{\k{a}}czyk, P.~Becker, M.~Bender, K.~Bennaceur,
  J.~Bonnard, Y.~Gao, A.~Idini, M.~Konieczka, M.~Kortelainen, L.~Pr{\'o}chniak,
  A.~M. Romero, W.~Satu{\l}a, Y.~Shi, T.~R. Werner, and L.~F. Yu.
\newblock ``Solution of universal nonrelativistic nuclear {DFT} equations in
  the cartesian deformed harmonic-oscillator basis. ({IX}) {HFODD} (v3.06h): a
  new version of the program''.
\newblock \href{https://dx.doi.org/10.1088/1361-6471/ac0a82}{Journal of Physics
  G: Nuclear and Particle Physics {\bf 48}, 102001}~(2021).

\bibitem{marevic2022hfbtho}
P.~Marevi{\'c}, N.~Schunck, E.~M. Ney, R.~Navarro P{\'e}rez, M.~Verri{\`e}re,
  and J.~O'Neal.
\newblock ``Axially-deformed solution of the
  {Skyrme--Hartree--Fock--Bogoliubov} equations using the transformed harmonic
  oscillator basis ({IV}) {HFBTHO} (v4.0): a new version of the program''.
\newblock \href{https://dx.doi.org/10.1016/j.cpc.2022.108367}{Computer Physics
  Communications {\bf 276}, 108367}~(2022).

\bibitem{lay2024fissionemulation}
Daniel Lay, Eric Flynn, Samuel~A. Giuliani, Witold Nazarewicz, and L{\'e}o
  Neufcourt.
\newblock ``Neural network emulation of spontaneous fission''.
\newblock \href{https://dx.doi.org/10.1103/PhysRevC.109.044305}{Physical Review
  C {\bf 109}, 044305}~(2024).

\bibitem{stratonovich1957method}
R.~L. Stratonovich.
\newblock ``A method for the computation of quantum distribution functions''.
\newblock Doklady Akademii Nauk SSSR {\bf 115}, 1097--1100~(1957).
\newblock  url:~\url{https://www.mathnet.ru/eng/dan22296}.

\bibitem{hubbard1959partition}
J.~Hubbard.
\newblock ``Calculation of partition functions''.
\newblock \href{https://dx.doi.org/10.1103/PhysRevLett.3.77}{Physical Review
  Letters {\bf 3}, 77--78}~(1959).

\bibitem{boyer1998tight}
Michel Boyer, Gilles Brassard, Peter H{\o}yer, and Alain Tapp.
\newblock ``Tight bounds on quantum searching''.
\newblock
  \href{https://dx.doi.org/10.1002/(SICI)1521-3978(199806)46:4/5<493::AID-PROP493>3.0.CO;2-P}{Fortschritte
  der Physik {\bf 46}, 493--505}~(1998).

\bibitem{durr1996minimum}
Christoph D{\"u}rr and Peter H{\o}yer.
\newblock ``A quantum algorithm for finding the minimum''~(1996).
\newblock
  \href{http://arxiv.org/abs/quant-ph/9607014}{arXiv:quant-ph/9607014}.

\bibitem{walesdoye1997basinhopping}
David~J. Wales and Jonathan P.~K. Doye.
\newblock ``Global optimization by basin-hopping and the lowest energy
  structures of {Lennard-Jones} clusters containing up to 110 atoms''.
\newblock \href{https://dx.doi.org/10.1021/jp970984n}{The Journal of Physical
  Chemistry A {\bf 101}, 5111--5116}~(1997).

\bibitem{jones1998ego}
Donald~R. Jones, Matthias Schonlau, and William~J. Welch.
\newblock ``Efficient global optimization of expensive black-box functions''.
\newblock \href{https://dx.doi.org/10.1023/A:1008306431147}{Journal of Global
  Optimization {\bf 13}, 455--492}~(1998).

\bibitem{dumitrescu2018nucleus}
E.~F. Dumitrescu, A.~J. McCaskey, G.~Hagen, G.~R. Jansen, T.~D. Morris,
  T.~Papenbrock, R.~C. Pooser, D.~J. Dean, and P.~Lougovski.
\newblock ``Cloud quantum computing of an atomic nucleus''.
\newblock \href{https://dx.doi.org/10.1103/PhysRevLett.120.210501}{Physical
  Review Letters {\bf 120}, 210501}~(2018).

\bibitem{stetcu2022variational}
I.~Stetcu, A.~Baroni, and J.~Carlson.
\newblock ``Variational approaches to constructing the many-body nuclear ground
  state for quantum computing''.
\newblock \href{https://dx.doi.org/10.1103/PhysRevC.105.064308}{Physical Review
  C {\bf 105}, 064308}~(2022).

\bibitem{kiss2022ordereducc}
Oriel Kiss, Michele Grossi, Pavel Lougovski, Federico Sanchez, Sofia
  Vallecorsa, and Thomas Papenbrock.
\newblock ``Quantum computing of the {$^{6}\mathrm{Li}$} nucleus via ordered
  unitary coupled clusters''.
\newblock \href{https://dx.doi.org/10.1103/PhysRevC.106.034325}{Physical Review
  C {\bf 106}, 034325}~(2022).

\bibitem{romero2022adapt}
A.~M. Romero, J.~Engel, Ho~Lun Tang, and Sophia~E. Economou.
\newblock ``Solving nuclear structure problems with the adaptive variational
  quantum algorithm''.
\newblock \href{https://dx.doi.org/10.1103/PhysRevC.105.064317}{Physical Review
  C {\bf 105}, 064317}~(2022).

\bibitem{stetcu2023projection}
I.~Stetcu, A.~Baroni, and J.~Carlson.
\newblock ``Projection algorithm for state preparation on quantum computers''.
\newblock \href{https://dx.doi.org/10.1103/PhysRevC.108.L031306}{Physical
  Review C {\bf 108}, L031306}~(2023).

\bibitem{liu2025pairingblockencoding}
Diyi Liu, Weijie Du, Lin Lin, James~P. Vary, and Chao Yang.
\newblock ``An efficient quantum circuit for block encoding a pairing
  {Hamiltonian}''.
\newblock \href{https://dx.doi.org/10.1016/j.jocs.2024.102480}{Journal of
  Computational Science {\bf 85}, 102480}~(2025).

\bibitem{costa2025annealing}
Emanuele Costa, Axel P{\'e}rez-Obiol, Javier Men{\'e}ndez, Arnau Rios, Artur
  Garc{\'i}a-S{\'a}ez, and Bruno Juli{\'a}-D{\'i}az.
\newblock ``A quantum annealing protocol to solve the nuclear shell model''.
\newblock \href{https://dx.doi.org/10.21468/SciPostPhys.19.2.062}{SciPost
  Physics {\bf 19}, 062}~(2025).

\bibitem{costa2026quasiparticle}
Emanuele Costa, Axel P{\'e}rez-Obiol, Javier Men{\'e}ndez, Arnau Rios, Artur
  Garc{\'i}a-S{\'a}ez, and Bruno Juli{\'a}-D{\'i}az.
\newblock ``Quasiparticle pairing encoding of atomic nuclei for quantum
  annealing''.
\newblock \href{https://dx.doi.org/10.1016/j.physletb.2025.140042}{Physics
  Letters B {\bf 872}, 140042}~(2026).

\bibitem{abhishek2026crankedzr}
Abhishek, Nabeel Salim, and P.~Arumugam.
\newblock ``Quantum simulation of cranked zirconium isotopes: A fixed-{$N$}
  approach with a structured number-conserving ansatz''~(2026).
\newblock  \href{http://arxiv.org/abs/2604.00844}{arXiv:2604.00844}.

\bibitem{baritompa2005grover}
W.~P. Baritompa, D.~W. Bulger, and G.~R. Wood.
\newblock ``Grover's quantum algorithm applied to global optimization''.
\newblock \href{https://dx.doi.org/10.1137/040605072}{SIAM Journal on
  Optimization {\bf 15}, 1170--1184}~(2005).

\bibitem{bulger2005basinhopping}
David Bulger.
\newblock ``Quantum basin hopping with gradient-based local
  optimisation''~(2005).
\newblock
  \href{http://arxiv.org/abs/quant-ph/0507193}{arXiv:quant-ph/0507193}.

\bibitem{zhu2009simulated}
Jing Zhu, Zhen Huang, and Sabre Kais.
\newblock ``Simulated quantum computation of global minima''.
\newblock \href{https://dx.doi.org/10.1080/00268970903117126}{Molecular Physics
  {\bf 107}, 2015--2023}~(2009).

\bibitem{soltaninia2025qgmf}
Mohammadreza Soltaninia and Junpeng Zhan.
\newblock ``Quantum global minimum finder based on variational quantum
  search''.
\newblock \href{https://dx.doi.org/10.1038/s41598-025-91407-z}{Scientific
  Reports {\bf 15}, 13880}~(2025).

\bibitem{cerveralierta2021metavqe}
Alba Cervera-Lierta, Jakob~S. Kottmann, and Al{\'a}n Aspuru-Guzik.
\newblock ``{Meta-Variational Quantum Eigensolver}: Learning energy profiles of
  parameterized {Hamiltonians} for quantum simulation''.
\newblock \href{https://dx.doi.org/10.1103/PRXQuantum.2.020329}{PRX Quantum
  {\bf 2}, 020329}~(2021).

\bibitem{delgado2021geometry}
Alain Delgado, Juan~Miguel Arrazola, Soran Jahangiri, Zeyue Niu, Josh Izaac,
  Chase Roberts, and Nathan Killoran.
\newblock ``Variational quantum algorithm for molecular geometry
  optimization''.
\newblock \href{https://dx.doi.org/10.1103/PhysRevA.104.052402}{Physical Review
  A {\bf 104}, 052402}~(2021).

\bibitem{yuan2021hamiltonianwavefunction}
Zhan-Hao Yuan, Tao Yin, and Dan-Bo Zhang.
\newblock ``Hybrid quantum-classical algorithms for solving quantum chemistry
  in {Hamiltonian--wave-function} space''.
\newblock \href{https://dx.doi.org/10.1103/PhysRevA.103.012413}{Physical Review
  A {\bf 103}, 012413}~(2021).

\bibitem{kassal2009geometry}
Ivan Kassal and Al{\'a}n Aspuru-Guzik.
\newblock ``Quantum algorithm for molecular properties and geometry
  optimization''.
\newblock \href{https://dx.doi.org/10.1063/1.3266959}{The Journal of Chemical
  Physics {\bf 131}, 224102}~(2009).

\bibitem{kosugi2023geometry}
Taichi Kosugi, Hirofumi Nishi, and Yu~ichiro Matsushita.
\newblock ``Exhaustive search for optimal molecular geometries using
  imaginary-time evolution on a quantum computer''.
\newblock \href{https://dx.doi.org/10.1038/s41534-023-00778-6}{npj Quantum
  Information {\bf 9}, 112}~(2023).

\bibitem{hirai2022electronsnuclei}
Hirotoshi Hirai, Takahiro Horiba, Soichi Shirai, Keita Kanno, Keita Omiya,
  Yuya~O. Nakagawa, and Sho Koh.
\newblock ``Molecular structure optimization based on electrons--nuclei quantum
  dynamics computation''.
\newblock \href{https://dx.doi.org/10.1021/acsomega.2c01546}{ACS Omega {\bf 7},
  19784--19793}~(2022).

\bibitem{arve1988staticpath}
P.~Arve, G.~Bertsch, B.~Lauritzen, and G.~Puddu.
\newblock ``Static path approximation for the nuclear partition function''.
\newblock \href{https://dx.doi.org/10.1016/0003-4916(88)90235-7}{Annals of
  Physics {\bf 183}, 309--319}~(1988).

\bibitem{koonin1997smmc}
S.~E. Koonin, D.~J. Dean, and K.~Langanke.
\newblock ``Shell model {Monte Carlo} methods''.
\newblock \href{https://dx.doi.org/10.1016/S0370-1573(96)00017-8}{Physics
  Reports {\bf 278}, 1--77}~(1997).

\bibitem{feynman1939forces}
R.~P. Feynman.
\newblock ``Forces in molecules''.
\newblock \href{https://dx.doi.org/10.1103/PhysRev.56.340}{Physical Review {\bf
  56}, 340--343}~(1939).

\bibitem{kramers1930paramagnetic}
H.~A. Kramers.
\newblock ``Th{\'e}orie g{\'e}n{\'e}rale de la rotation paramagn{\'e}tique dans
  les cristaux''.
\newblock Proceedings Koninklijke Akademie van Wetenschappen {\bf 33},
  959--972~(1930).
\newblock  url:~\url{https://dwc.knaw.nl/DL/publications/PU00015981.pdf}.

\bibitem{jordanwigner1928mapping}
P.~Jordan and E.~Wigner.
\newblock ``{\"U}ber das {Paulische} {\"a}quivalenzverbot''.
\newblock \href{https://dx.doi.org/10.1007/BF01331938}{Zeitschrift f{\"u}r
  Physik {\bf 47}, 631--651}~(1928).

\bibitem{barangerkumar1968pairingquadrupole}
Michel Baranger and Krishna Kumar.
\newblock ``Nuclear deformations in the pairing-plus-quadrupole model: ({II}).
  discussion of validity of the model''.
\newblock \href{https://dx.doi.org/10.1016/0375-9474(68)90370-9}{Nuclear
  Physics A {\bf 110}, 490--528}~(1968).

\bibitem{cleve1998algorithms}
Richard Cleve, Artur Ekert, Chiara Macchiavello, and Michele Mosca.
\newblock ``Quantum algorithms revisited''.
\newblock \href{https://dx.doi.org/10.1098/rspa.1998.0164}{Proceedings of the
  Royal Society of London. Series A: Mathematical, Physical and Engineering
  Sciences {\bf 454}, 339--354}~(1998).

\bibitem{abrams1999eigenvalues}
Daniel~S. Abrams and Seth Lloyd.
\newblock ``Quantum algorithm providing exponential speed increase for finding
  eigenvalues and eigenvectors''.
\newblock \href{https://dx.doi.org/10.1103/PhysRevLett.83.5162}{Physical Review
  Letters {\bf 83}, 5162--5165}~(1999).

\bibitem{poulinwocjan2009ground}
David Poulin and Pawel Wocjan.
\newblock ``Preparing ground states of quantum many-body systems on a quantum
  computer''.
\newblock \href{https://dx.doi.org/10.1103/PhysRevLett.102.130503}{Physical
  Review Letters {\bf 102}, 130503}~(2009).

\bibitem{lintong2020filter}
Lin Lin and Yu~Tong.
\newblock ``Optimal polynomial based quantum eigenstate filtering with
  application to solving quantum linear systems''.
\newblock \href{https://dx.doi.org/10.22331/q-2020-11-11-361}{Quantum {\bf 4},
  361}~(2020).

\bibitem{lintong2020ground}
Lin Lin and Yu~Tong.
\newblock ``Near-optimal ground state preparation''.
\newblock \href{https://dx.doi.org/10.22331/q-2020-12-14-372}{Quantum {\bf 4},
  372}~(2020).

\bibitem{jansen2007adiabatic}
Sabine Jansen, Mary-Beth Ruskai, and Ruedi Seiler.
\newblock ``Bounds for the adiabatic approximation with applications to quantum
  computation''.
\newblock \href{https://dx.doi.org/10.1063/1.2798382}{Journal of Mathematical
  Physics {\bf 48}, 102111}~(2007).

\bibitem{trotter1959product}
H.~F. Trotter.
\newblock ``On the product of semi-groups of operators''.
\newblock
  \href{https://dx.doi.org/10.1090/S0002-9939-1959-0108732-6}{Proceedings of
  the American Mathematical Society {\bf 10}, 545--551}~(1959).

\bibitem{suzuki1990fractal}
Masuo Suzuki.
\newblock ``Fractal decomposition of exponential operators with applications to
  many-body theories and {Monte Carlo} simulations''.
\newblock \href{https://dx.doi.org/10.1016/0375-9601(90)90962-N}{Physics
  Letters A {\bf 146}, 319--323}~(1990).

\bibitem{bennett1973reversibility}
C.~H. Bennett.
\newblock ``Logical reversibility of computation''.
\newblock \href{https://dx.doi.org/10.1147/rd.176.0525}{IBM Journal of Research
  and Development {\bf 17}, 525--532}~(1973).

\bibitem{gershgorin1931bounds}
S.~Ger{\v{s}}gorin.
\newblock ``{\"U}ber die abgrenzung der eigenwerte einer matrix''.
\newblock Bulletin de l'Acad{\'e}mie des Sciences de l'URSS. Classe des
  sciences math{\'e}matiques et naturellesPages 749--754~(1931).
\newblock  url:~\url{https://www.mathnet.ru/eng/im5235}.

\bibitem{hornjohnson2012matrix}
Roger~A. Horn and Charles~R. Johnson.
\newblock ``Matrix analysis''.
\newblock \href{https://dx.doi.org/10.1017/CBO9781139020411}{Cambridge
  University Press}. ~(2012).
\newblock 2 edition.

\bibitem{lintong2022heisenberg}
Lin Lin and Yu~Tong.
\newblock ``{Heisenberg-Limited Ground-State Energy Estimation} for early
  fault-tolerant quantum computers''.
\newblock \href{https://dx.doi.org/10.1103/PRXQuantum.3.010318}{PRX Quantum
  {\bf 3}, 010318}~(2022).

\bibitem{patel2026taperedqpe}
Dhrumil Patel, Shi Jie~Samuel Tan, Yi{\u{g}}it Suba{\c{s}}{\i}, and Andrew~T.
  Sornborger.
\newblock ``Optimal coherent quantum phase estimation via tapering''.
\newblock \href{https://dx.doi.org/10.1103/l5y6-6zxv}{PRX Quantum {\bf 7},
  020302}~(2026).

\bibitem{gunther2026randomizedqpe}
Jakob G{\"u}nther, Freek Witteveen, Alexander Schmidhuber, Marek Miller,
  Matthias Christandl, and Aram~W. Harrow.
\newblock ``Phase estimation with partially randomized time evolution''.
\newblock \href{https://dx.doi.org/10.1103/ynxb-p2xq}{PRX Quantum {\bf 7},
  020332}~(2026).

\bibitem{childs2021trotter}
Andrew~M. Childs, Yuan Su, Minh~C. Tran, Nathan Wiebe, and Shuchen Zhu.
\newblock ``Theory of {Trotter} error with commutator scaling''.
\newblock \href{https://dx.doi.org/10.1103/PhysRevX.11.011020}{Physical Review
  X {\bf 11}, 011020}~(2021).

\bibitem{lowchuang2017qsp}
Guang~Hao Low and Isaac~L. Chuang.
\newblock ``Optimal {Hamiltonian} simulation by quantum signal processing''.
\newblock \href{https://dx.doi.org/10.1103/PhysRevLett.118.010501}{Physical
  Review Letters {\bf 118}, 010501}~(2017).

\bibitem{lowchuang2019qubitization}
Guang~Hao Low and Isaac~L. Chuang.
\newblock ``{Hamiltonian Simulation by Qubitization}''.
\newblock \href{https://dx.doi.org/10.22331/q-2019-07-12-163}{Quantum {\bf 3},
  163}~(2019).

\bibitem{yoder2014fixedpoint}
Theodore~J. Yoder, Guang~Hao Low, and Isaac~L. Chuang.
\newblock ``Fixed-point quantum search with an optimal number of queries''.
\newblock \href{https://dx.doi.org/10.1103/PhysRevLett.113.210501}{Physical
  Review Letters {\bf 113}, 210501}~(2014).
\newblock  \href{http://arxiv.org/abs/1409.3305}{arXiv:1409.3305}.

\bibitem{rendon2022tapering}
Gumaro Rendon, Taku Izubuchi, and Yuta Kikuchi.
\newblock ``Effects of cosine tapering window on quantum phase estimation''.
\newblock \href{https://dx.doi.org/10.1103/PhysRevD.106.034503}{Physical Review
  D {\bf 106}, 034503}~(2022).
\newblock  \href{http://arxiv.org/abs/2110.09590}{arXiv:2110.09590}.

\bibitem{caurier2005shellmodel}
E.~Caurier, G.~Mart{\'i}nez-Pinedo, F.~Nowacki, A.~Poves, and A.~P. Zuker.
\newblock ``The shell model as a unified view of nuclear structure''.
\newblock \href{https://dx.doi.org/10.1103/RevModPhys.77.427}{Reviews of Modern
  Physics {\bf 77}, 427--488}~(2005).

\bibitem{grover1996search}
Lov~K. Grover.
\newblock ``A fast quantum mechanical algorithm for database search''.
\newblock In Proceedings of the Twenty-Eighth Annual ACM Symposium on Theory of
  Computing.
\newblock \href{https://dx.doi.org/10.1145/237814.237866}{Pages 212--219}.
\newblock Association for Computing Machinery~(1996).

\bibitem{brassard2002amplitude}
Gilles Brassard, Peter H{\o}yer, Michele Mosca, and Alain Tapp.
\newblock ``Quantum amplitude amplification and estimation''.
\newblock In Quantum Computation and Information.
\newblock \href{https://dx.doi.org/10.1090/conm/305/05215}{Volume 305 of
  Contemporary Mathematics, pages 53--74}.
\newblock American Mathematical Society~(2002).

\bibitem{hoyer2003boundederror}
Peter H{\o}yer, Michele Mosca, and Ronald de~Wolf.
\newblock ``Quantum search on bounded-error inputs''.
\newblock In Automata, Languages and Programming.
\newblock \href{https://dx.doi.org/10.1007/3-540-45061-0_25}{Volume 2719 of
  Lecture Notes in Computer Science, pages 291--299}.
\newblock Springer~(2003).

\bibitem{neldermead1965simplex}
J.~A. Nelder and R.~Mead.
\newblock ``A simplex method for function minimization''.
\newblock \href{https://dx.doi.org/10.1093/comjnl/7.4.308}{The Computer Journal
  {\bf 7}, 308--313}~(1965).

\bibitem{stornprice1997de}
Rainer Storn and Kenneth Price.
\newblock ``Differential evolution---a simple and efficient heuristic for
  global optimization over continuous spaces''.
\newblock \href{https://dx.doi.org/10.1023/A:1008202821328}{Journal of Global
  Optimization {\bf 11}, 341--359}~(1997).

\bibitem{brown2006usd}
B.~Alex Brown and W.~A. Richter.
\newblock ``New ``{USD}'' hamiltonians for the \textit{sd} shell''.
\newblock \href{https://dx.doi.org/10.1103/PhysRevC.74.034315}{Physical Review
  C {\bf 74}, 034315}~(2006).

\bibitem{jokiniemi2023muoncapture}
L.~Jokiniemi, T.~Miyagi, S.~R. Stroberg, J.~D. Holt, J.~Kotila, and J.~Suhonen.
\newblock ``Ab initio calculation of muon capture on $^{24}${Mg}''.
\newblock \href{https://dx.doi.org/10.1103/PhysRevC.107.014327}{Physical Review
  C {\bf 107}, 014327}~(2023).

\bibitem{ambainis2010variabletime}
Andris Ambainis.
\newblock ``Quantum search with variable times''.
\newblock \href{https://dx.doi.org/10.1007/s00224-009-9219-1}{Theory of
  Computing Systems {\bf 47}, 786--807}~(2010).
\newblock
  \href{http://arxiv.org/abs/quant-ph/0609168}{arXiv:quant-ph/0609168}.

\bibitem{barrett2013nocore}
Bruce~R. Barrett, Petr Navr{\'a}til, and James~P. Vary.
\newblock ``Ab initio no core shell model''.
\newblock \href{https://dx.doi.org/10.1016/j.ppnp.2012.10.003}{Progress in
  Particle and Nuclear Physics {\bf 69}, 131--181}~(2013).

\bibitem{lee2023evaluating}
Seunghoon Lee, Joonho Lee, Huanchen Zhai, Yu~Tong, Alexander~M. Dalzell,
  Ashutosh Kumar, Phillip Helms, Johnnie Gray, Zhi-Hao Cui, Wenyuan Liu,
  Michael Kastoryano, Ryan Babbush, John Preskill, David~R. Reichman, Earl~T.
  Campbell, Edward~F. Valeev, Lin Lin, and Garnet Kin-Lic Chan.
\newblock ``Evaluating the evidence for exponential quantum advantage in
  ground-state quantum chemistry''.
\newblock \href{https://dx.doi.org/10.1038/s41467-023-37587-6}{Nature
  Communications {\bf 14}, 1952}~(2023).

\end{thebibliography}

\appendix

\section{Numerical checks of the schematic model}
\label{app:checks}

These checks support the circuit-level calculations of
Sec.~\ref{sec:validation}. Exact eigenstates and spectra enter them only
after a search result has been fixed.

\paragraph{Field--response balance.}
Continuous minimization gives \(q=-0.73067649\) and
\(\langle Q\rangle=-1.91802577\). The two sides of
Eq.~\eqref{eq:stationary-field-balance} are \(-1.53442064\) and
\(-1.53442062\). This confirms the stationary balance; it does not change the
discrete minimum selected by the search.

\paragraph{Preparation.}
The path gaps in Table~\ref{tab:prep-extension} are sampled at 101 points
along each path; the smallest is \(1.2\). On the four-point grid the
coordinate leakage is below \(9.0\times10^{-14}\), and preparation followed
by its inverse returns the input to statevector precision. In the original
two-coordinate run the leakage is below \(1.7\times10^{-13}\) and the
returned branch has ground-state fidelity \(0.906617\).

\paragraph{Spectral and evolution checks.}
The four-point QPE with nine phase qubits has bin width \(0.0148005\). Two
Suzuki repetitions per base power give 1022 repetitions in one forward QPE;
the base-power step has spectral-norm error \(0.04917\) and process
infidelity \(2.57\times10^{-4}\). All 24 eigenvalues of the four
fixed-particle-number blocks map to phases in \([0.07916,0.79481]\), inside
the certified interval. In the eight-point extension, the 16-qubit forward
circuit at \(q=-0.72\) gives decoded energy \(-2.5969048\), within one bin of
the exact ground energy. For evolution time \(t=0.2\) the second-order Suzuki
error falls from \(4.98\times10^{-3}\) with one repetition to
\(1.22\times10^{-3}\), \(3.03\times10^{-4}\), and \(7.57\times10^{-5}\) with
two, four, and eight (Fig.~\ref{fig:structured-validation}); no mixing of
coordinate blocks is observed. A separate five-bit QPE test gives
\(\widetilde E=-2.5625\) for the reference energy \(-2.5006544\), within half
the bin width \(0.21875\), with inverse-QPE residual \(3.59\times10^{-12}\).

For the \(4\times4\) two-coordinate grid, the fixed-particle-number
Hamiltonian has dimension \(16\times6=96\). Direct and coherent Hamiltonians
agree block by block to \(1.78\times10^{-15}\). With ten phase qubits the bin
width is \(0.0175089\), and the original run's returned energy estimate,
\(-2.7571786\), lies \(0.0178835\) above the exact minimum.

\paragraph{Search separation and execution counts.}
The search routine neither imports nor calls the exact-diagonalization
module; an integration test disables every eigensolver in the repository,
and the routine writes its returned coordinate and energy estimate to disk
before exact validation is allowed. In the four-point trajectory, the
initial modal bin 43 decodes to \(-2.5731004\) and the accepted bin 42 to
\(-2.5879010\). The single amplifying round gives coordinate probabilities
\((0.03146,0.92102,0.02029,0.02723)\) with clean-work probability
\(0.90252\). Each preparation or inverse preparation uses eight ASP slices,
so the 569 preparation calls of the one-dimensional run correspond to 4552
ASP evolution calls. The two-coordinate QPE stage decomposes to
\(3.0\times10^6\) instructions at depth \(2.5\times10^6\) in the basis of
Table~\ref{tab:prep-extension}.

\section{Preparation schedules}
\label{app:schedules}

Table~\ref{tab:schedules} gives the worst and mean branch fidelity of the
two-coordinate preparation for three ramps and several total times,
simulated with the actual product-formula circuits. With time steps of
\(0.5\) every ramp levels off between \(0.93\) and \(0.96\): the
product-formula error, not the ramp, then limits the fidelity. With steps of
\(0.25\) the smooth ramps pass \(0.98\) at \(T=12\), while the linear ramp
needs more than \(T=20\).

\begin{table*}[tbp]
\caption{Worst (mean) branch fidelity of the \(4\times4\) two-coordinate
preparation. Linear: \(s=u\); smoothstep: \(s=3u^2-2u^3\); smootherstep:
\(s=10u^3-15u^4+6u^5\); one second-order Suzuki step per time step.}
\label{tab:schedules}
\begin{ruledtabular}
\begin{tabular}{ccccc}
step & \(T\) & linear & smoothstep & smootherstep \\
\hline
0.5 & 4 & 0.744 (0.921) & 0.725 (0.917) & 0.707 (0.908) \\
0.5 & 12 & 0.924 (0.984) & 0.937 (0.984) & 0.933 (0.982) \\
0.5 & 20 & 0.934 (0.984) & 0.949 (0.987) & 0.959 (0.987) \\
0.25 & 8 & 0.943 (0.980) & 0.961 (0.987) & 0.961 (0.985) \\
0.25 & 12 & 0.969 (0.992) & 0.988 (0.997) & 0.984 (0.997) \\
0.25 & 16 & 0.978 (0.994) & 0.993 (0.999) & 0.995 (0.999) \\
0.25 & 20 & 0.984 (0.997) & 0.996 (0.999) & 0.997 (0.999) \\
\end{tabular}
\end{ruledtabular}
\end{table*}

\section{Phase-estimation precision and windows}
\label{app:precision}

Table~\ref{tab:precision} shows how the phase window controls the two
quantities that the threshold test depends on. ``Marked'' is the smallest
weight that a branch below the threshold places below it, and ``leakage''
the largest weight that a branch above the threshold places below it, over
all thresholds at least two bins from every grid energy, with the
smoothstep preparation (\(F_{\min}=0.988\)). A tapered window keeps the
marked weight at the preparation fidelity and cuts the leakage by three
orders of magnitude. With eight repetitions per base power, the largest
shift between a branch's modal QPE energy and its exact ground energy is
\(0.010\) for ten phase qubits, \(0.005\) for eleven, and \(0.004\) for
twelve; with two repetitions it is \(0.036\).

\begin{table}[tbp]
\caption{Threshold-test quality of the \(4\times4\) two-coordinate QPE with
eight repetitions per base power, for thresholds at least two bins from
every grid energy.}
\label{tab:precision}
\begin{ruledtabular}
\begin{tabular}{clcc}
\(m\) & window & marked & leakage \\
\hline
11 & uniform & 0.945 & \(5.0\times10^{-2}\) \\
11 & sine & 0.988 & \(1.2\times10^{-3}\) \\
11 & Kaiser(6) & 0.988 & \(2.9\times10^{-5}\) \\
11 & Kaiser(10) & 0.982 & \(2.8\times10^{-3}\) \\
12 & uniform & 0.945 & \(4.8\times10^{-2}\) \\
12 & sine & 0.987 & \(1.1\times10^{-3}\) \\
12 & Kaiser(6) & 0.988 & \(2.6\times10^{-5}\) \\
12 & Kaiser(10) & 0.981 & \(5.3\times10^{-4}\) \\
\end{tabular}
\end{ruledtabular}
\end{table}

\section{Fixed-point sequence and simulation method}
\label{app:simulation}

For \(\delta=0.1\), sequences of length \(L=3\), \(5\), and \(7\) guarantee
\(P_L\geq0.99\) for marked weights above \(0.579\), \(0.287\), and \(0.163\),
respectively. The searches of Table~\ref{tab:robust-search} declare a floor
of \(0.5\) and therefore use \(L=5\). The phases follow
Ref.~\cite{yoder2014fixedpoint}; a two-level simulation reproduces
Eq.~\eqref{eq:fixed-point-success} to \(10^{-12}\), and the circuit
reproduces it on a toy model with ground-state weight \(0.6\).

Simulating the searches gate by gate is expensive because every phase bit
repeats the same controlled product-formula step many times. We therefore
simulate the same circuits by their operators: the unitary of one
controlled step is computed from the synthesized gates and raised to the
required powers, and the preparation, window, inverse Fourier transform, and
diffusion are applied as the operators of their circuits. No eigenvector,
eigenvalue, or exact exponential enters; product-formula and
phase-estimation errors are kept exactly. On a reduced instance the result
agrees with gate-level statevector simulation to \(10^{-8}\) for
preparation and QPE, both threshold oracles, and two-iteration searches, and
seeded searches follow identical trajectories. The method changes the cost of
simulation, not any logical resource count.

\section{Shell-model validation}
\label{app:sd-validation}

Table~\ref{tab:sd-validation} lists the USDB spectroscopy computed with the
implementation used for the landscapes of Sec.~\ref{sec:sd-shell}, with
effective charges \(e_p=1.5\), \(e_n=0.5\) and oscillator energy
\(\hbar\omega=45A^{-1/3}-25A^{-2/3}\) MeV. The interaction file is the
public USDB file in KSHELL format.

\begin{table*}[tbp]
\caption{USDB ground energies \(E_0\) and first \(2^+\) excitation energies
(MeV), \(B(E2;2^+_1\to0^+)\) (e\(^2\)fm\(^4\)), and spectroscopic
quadrupole moments \(Q(2^+_1)\) (e\,fm\(^2\)). ``Dim.'' is the dimension of
the proton--neutron product space of determinants.}
\label{tab:sd-validation}
\begin{ruledtabular}
\begin{tabular}{lrrrrr}
 & Dim. & \(E_0\) & \(E(2^+_1)\) & \(B(E2)\) & \(Q(2^+_1)\) \\
\hline
\(^{20}\)Ne & 4\,356 & \(-40.472\) & 1.747 & 60.1 & \(-15.8\) \\
\(^{24}\)Mg & 245\,025 & \(-87.104\) & 1.502 & 95.4 & \(-19.3\) \\
\(^{28}\)Si & 853\,776 & \(-135.861\) & 1.932 & 99.8 & \(+20.9\) \\
\end{tabular}
\end{ruledtabular}
\end{table*}

\end{document}